\documentclass{layoutTechReport/egpubl}
\usepackage{layoutTechReport/egsr2021}
 
\SpecialIssuePaper         % uncomment for final version of Computer Graphics
\SpecialIssueSubmission    % uncomment for submission to , special issue
 
\usepackage[T1]{fontenc}
\usepackage{layoutTechReport/dfadobe}

\biberVersion
\BibtexOrBiblatex
\usepackage[backend=biber,bibstyle=layoutTechReport/EG,citestyle=alphabetic,backref=true,useprefix=true]{biblatex} 
\electronicVersion
\PrintedOrElectronic

\ifpdf \usepackage[pdftex]{graphicx} \pdfcompresslevel=9
\else \usepackage[dvips]{graphicx} \fi

\usepackage{layoutTechReport/egweblnk}

\usepackage{amsmath}
\usepackage{amsfonts} 
\usepackage{graphicx}
\usepackage{color}
\usepackage{xcolor}
\usepackage{subcaption}
\usepackage{float}
\usepackage{tabu}                      
\usepackage{booktabs}                  
\usepackage{multirow}
\usepackage{listings}
\usepackage{todonotes}
\usepackage{textcomp}
\usepackage{lmodern}
\usepackage{makecell}
\usepackage{colortbl}

\DeclareUnicodeCharacter{8680}{\rarrow}

\definecolor{codegreen}{rgb}{0,0.6,0}
\definecolor{codegray}{rgb}{0.5,0.5,0.5}
\definecolor{codepurple}{rgb}{0.58,0,0.82}
\definecolor{backcolour}{rgb}{0.95,0.95,0.92}

\lstdefinestyle{mystyle}{
    backgroundcolor=\color{backcolour},   
    commentstyle=\color{codegreen},
    keywordstyle=\color{magenta},
    numberstyle=\tiny\color{codegray},
    stringstyle=\color{codepurple},
    basicstyle=\ttfamily\small,
    breakatwhitespace=false,         
    breaklines=true,                 
    captionpos=b,                    
    keepspaces=true,                 
    numbers=left,                    
    numbersep=5pt,                  
    showspaces=false,                
    showstringspaces=false,
    showtabs=false,                  
    tabsize=2
}

\newcommand{\enum}[1]{\texttt{#1}}
\newcommand{\code}[1]{\textit{#1}}

\title[Rendering Point Clouds with Compute Shaders]{Rendering Point Clouds with Compute Shaders and Vertex Order Optimization}

\author[Markus Sch{\"u}tz \& Bernhard Kerbl \& Michael Wimmer]
{\parbox{\textwidth}{\centering Markus Sch{\"u}tz,
        Bernhard Kerbl,
        Michael Wimmer 
        }
        \\
{
    \parbox{\textwidth}{
        \centering TU Wien, Institute of Visual Computing
    }
}
}

\begin{document}

\teaser{
 \vspace{-0.5cm}
 \includegraphics[width=\textwidth]{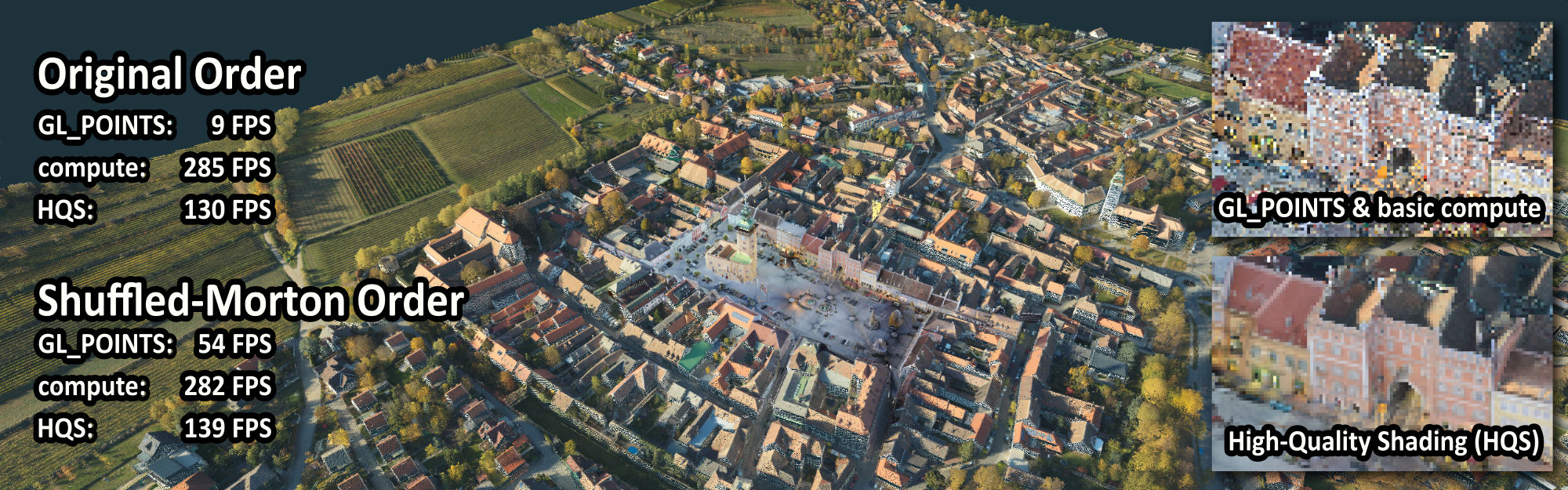}
        \caption{Performance of \enum{GL\_POINTS} compared to a  compute shader that performs local reduction and and early-z testing, and a high-quality compute shader that blends overlapping points. Retz point cloud (145 million points) courtesy of Riegl.}
    \label{fig:teaser}
}

\maketitle
%-------------------------------------------------------------------------

\begin{abstract}

While commodity GPUs provide a continuously growing range of features and sophisticated methods for accelerating compute jobs, many state-of-the-art solutions for point cloud rendering still rely on the provided point primitives (\enum{GL\_POINTS}, \enum{POINTLIST}, ...) of graphics APIs for image synthesis. In this paper, we present several compute-based point cloud rendering approaches that outperform the hardware pipeline by up to an order of magnitude and achieve significantly better frame times than previous compute-based methods. Beyond basic closest-point rendering, we also introduce a fast, high-quality variant to reduce aliasing. We present and evaluate several variants of our proposed methods with different flavors of optimization, in order to ensure their applicability and achieve optimal performance on a range of platforms and architectures with varying support for novel GPU hardware features. During our experiments, the observed peak performance was reached rendering 796 million points (12.7GB) at rates of 62 to 64 frames per second (50 billion points per second, 802GB/s) on an RTX 3090 without the use of level-of-detail structures.

We further introduce an optimized vertex order for point clouds to boost the efficiency of GL\_POINTS by a factor of 5$\times$ in cases where hardware rendering is compulsory. We compare different orderings and show that Morton sorted buffers are faster for some viewpoints, while shuffled vertex buffers are faster in others. In contrast, combining both approaches by first sorting according to Morton-code and shuffling the resulting sequence in batches of 128 points leads to a vertex buffer layout with high rendering performance and low sensitivity to viewpoint changes.

\end{abstract}

\section{Introduction}

Point clouds are three-dimensional models that consist of individual points with no connectivity. They are typically obtained by reconstructing the real world, for example with laser scanners or photogrammetry. Laser scanners obtain point-sampled surface models by measuring the distance from the scanner to surrounding surfaces, and then transforming the distance values and the known orientation of the scanner into 3D coordinates. Photogrammetry uses multiple photos to create a 3D model that best fits these images. The amount of points produced by these methods ranges from a few million up to trillions of points, depending on the scan resolution and the extent of the scanned area. The \textit{Actueel Hoogtebestand Nederland} (AHN2)~\cite{AHN2} data set, for example, is an aerial laser scan of the entire Netherlands, comprising 640 billion points, for a total of 1.6 terabytes in compressed form~\cite{MartinezRubi2015}. Terrestrial laser scans (e.g., of buildings, monuments, caves, smaller regions) usually yield from tens of millions up to a few billion points. 

The rendering of large point clouds is an active field of research that includes topics such as high-quality rendering, as well as the generation and effective use of level-of-detail (LOD) structures. Arguably, the most widespread solution for rendering a given set of points is by using the native point primitive that modern graphics APIs provide in addition to lines and triangles, such as \enum{GL\_POINTS} for OpenGL and WebGL, \enum{D3D11\_PRIMITIVE\_TOPOLOGY\_POINTLIST} for DirectX, \enum{VK\_PRIMITIVE\_TOPOLOGY\_POINT\_LIST} for Vulkan and ``point-list" for WebGPU. A key property of point primitives is that they require only one single coordinate vector as input, thereby keeping vertex buffer usage and bandwidth requirements to a minimum. OpenGL and Vulkan enable developers to specify the pixel size of points inside the vertex shader (rasterizing them as rectangles), whereas DirectX only supports a size of 1 pixel per point. Rendering larger points in DirectX and DirectX-based solutions (e.g., WebGL, WebGPU) requires developers to emulate sized points. In case of WebGL, this is done by instancing a quad model at each point location, as seen in the backend ANGLE~\cite{ANGLE_Instancing}. 
\enum{GL\_POINTS} is the standard high-performance point rendering primitive in research (e.g.,~\cite{Richter2015,RichterThesis,morel2017android,LPC,RayCaching2019,SAINZ2004869,VAST:VAST04:105-114,DynamicStackings2020}) and software (e.g., CloudCompare~\cite{CloudCompare}, Potree~\cite{Potree}, QGIS~\cite{QGIS}, Cesium~\cite{Cesium}). It is also used in high-quality techniques that rely on quads with specific pixel sizes~\cite{guennebaud:2003,scheiblauer-2011-chnt}, as well as impostors~\cite{DynamicStackings2020} and particle systems~\cite{LODParticleSystem}. However, in this paper, we will only consider cases where each point is drawn to at most one pixel of the framebuffer.

Modern graphics processing units (GPUs) are capable of rendering several million points (smartphones, integrated graphics) up to around 100 million points (high-end GPUs) in real-time (60 fps) using \enum{GL\_POINTS} or its counterparts. Rendering arbitrarily large point clouds, however, requires hierarchical level-of-detail (LOD) structures that only load and render a small subset of the full model at any given time. Unfortunately, generating these structures requires time-consuming preprocessing steps that impede quick inspection of larger scenes. In this paper, we focus on raising the raw rendering performance of given point cloud data sets by exploiting the GPU compute pipeline, as well as low-overhead vertex reordering schemes. 
Based on our compute-based variants, we also propose an easy-to-implement, high-quality point cloud rendering method that can reduce aliasing while achieving higher frame rates than the hardware pipeline.
Since different hardware and graphics APIs impose diverse restrictions on the features that are exposed by compute shaders, we explore several variations of our compute-based solutions that exploit different feature levels to heed these limitations.
%allowing developers to pick the right one for the targeted system. 
For instance, developers that target the soon-to-be-released WebGPU standard may %will be able %to access atomic operations 
adopt our basic variants, while users of lower-level graphics APIs can exploit hardware-accelerated primitives for peak performance. 
%is currently under development and likely to include 32 bit atomicMin that suffices to implement the high-quality shader (using two geometry passes). Faster variations with a single geometry pass require 64 bit atomicMin and subgroup communication, however, which are discussed as extensions but it is currently not certain if and to what extend they will become available. 
%We also implement an anti-aliasing solution that blends overlapping fragments together to achieve high-quality splatting. Unlike other approaches, we do not attempt to correctly fill gaps with the help of surface-reconstruction methods, and instead refer to Section~\ref{sec:hole_filling} for a list of methods that can be used to fill holes in post-process. 
Hence, our contributions to the state of the art are:

\begin{itemize}
  \item Suitable alternatives to \textit{glDrawArrays(}\enum{GL\_POINTS}\textit{, ...)}, based on compute shaders that use 64bit atomic operations to draw into an interleaved depth and color buffer. 
  \item Performance improvements and extensions to the basic compute shader with early-z (as suggested in ~\cite{Gnther2013AGP}) and group-wide reduction of points with warp-level primitives to reduce contention in global GPU memory.
  \item A high-quality shader that provides anti-aliasing within pixels by blending overlapping points together (similar to mipmapping), while still being faster than the aliased results of \enum{GL\_POINTS}. The basic version, \textit{HQS}, is a simplified and compute-based implementation of the blending algorithm in \cite{Botsch:HQS}, while the \textit{HQS1R} version reduces memory accesses by updating the sum of colors and fragment counts with a single atomic instruction per point.
  \item An easy-to-implement policy for rearranging points in a shuffled Morton order that significantly improves stability and rendering performance with \enum{GL\_POINTS}.
  \item A thorough evaluation to quantify the impact of using different compute shader techniques, point orderings and GPUs for rendering a range of point-cloud data sets, both unstructured and embedded in LOD data structures.
\end{itemize}

% List non-contributions? E.g.:

% \begin{itemize}
%   \item Although we introduce shuffled-Morton ordering with a batch-size of 128 points as a major contribution, we currently can not claim that this is the most efficient ordering. 
% \end{itemize}

\section{Related Work}

\subsection{Compute-based Triangle Rasterization}

Although the GPU's rendering pipeline is becoming increasingly programmable, its implementation in hardware necessitates numerous restrictions. The general-purpose capabilities of modern GPUs have led developers to pursue parallel software rasterizers as an alternative. 
Freepipe~\cite{10.1145/1730804.1730817} describes one of the first practical software implementations running on a GPU, using a single thread per triangle. 
The authors report that Freepipe surpasses the performance of OpenGL-based solutions in carefully selected applications.
Later approaches managed to achieve a higher level of parallelism and significantly improved performance via tile-based rendering and coverage masks~\cite{10.1145/2766973, 10.1145/2018323.2018337}. 
Complete sort-middle streaming pipelines have been realized in both CUDA and OpenCL~\cite{Kenzel:2018, Kim2021}.
However, none of the solutions that have been presented to date can reliably outperform the hardware pipeline in the average case. 

\subsection{Compute-based Point Cloud Rendering}

Günther et al.~\cite{Gnther2013AGP} were the first to suggest compute-based point rendering as an alternative to the standard point primitives of OpenGL. Their OpenCL implementation uses a busy-loop to wait until a pixel is free to write to, locks the pixel with an atomic compare-and-swap, updates the depth and color buffers, and afterward unlocks the pixel again. The authors observed that adding a custom early-z test enables the compute-based approach to scale significantly better with the number of fragments per pixel than \enum{GL\_POINTS}. 
Lukac et al.~\cite{HybrodPointVolume} use a very similar \code{atomicMin}-based algorithm in their hybrid point and volume renderer. However, their method is not thread-safe and leads to artifacts with dense point clouds.
Instead of a busy-loop, our method updates an interleaved depth and color buffer with a single atomic instruction. 
Marrs et al.~\cite{Marrs2018Shadows} use compute-based point-cloud rendering as a means to reproject a depth map to several different views. Since no colors are required, they simply use \code{InterlockedMax} (DirectX's counterpart to \code{atomicMax}) to write the largest depth into the target buffer.
In contrast, our approach writes depth and color information and exploits local reduction to maximize performance.

\subsection{Hole Filling}
\label{sec:hole_filling}

The approaches presented in this paper only address rendering at most one pixel per point, which may lead to holes if the point density is not high enough from a given view of the scene. However, several hole-filling algorithms have been proposed to eliminate the gaps between points of a point cloud---many of them via fast screen-space methods that operate by analyzing the neighborhood of a pixel.

Grossman and Dally \cite{grossman1998point}, Pintus et al.~\cite{Pintus2011} and Marroquim et al.~\cite{Marroquim:2007:pbg} first draw points and then build an image pyramid (e.g., \cite{strengert2006pyramid}) that is subsequently used to fill gaps. This involves determining hidden surfaces that were rendered through gaps between points of the front-most visible surface. Afterward, background pixels and points that are determined to be hidden are replaced with colors from the image pyramid. Instead of an image pyramid, Rosenthal and Linsen~\cite{Rosenthal2008} use an iterative approach that covers wider areas with higher iteration counts, but often already succeeds with fewer iterations. In contrast to methods based on computing color values between the gaps of rendered pixels, Auto Splats~\cite{AutoSplats} takes an additional step to compute oriented and sized splats, based on the initial rendering, and draws these splats in a separate rendering pass. The resulting splats contain normals that can be used for shading, and their size is adapted to fill the gaps.

\subsection{High-Quality Rendering}

In addition to visible holes, rendering point clouds is prone to aliasing artifacts due to the lack of mipmapping. Arikan et al. \cite{arikan-2015-dmrt} create multiple textured depth maps out of point clouds, and then ray-trace the depth maps in order to obtain a high-quality reconstruction of the underlying point cloud data.
Surface splatting~\cite{zwicker2001surface} is an extension of point clouds that uses oriented disks as a better representation of the underlying surface. The splat sizes are adjusted to cover holes, and overlapping splats are blended together, which results in anti-aliasing similar to mipmapping. Botsch et al.~\cite{Botsch:HQS} implement an efficient surface-splatting method that uses a depth pass to generate a shifted depth buffer, an attribute pass to sum up weighted fragments that pass the depth test, and a normalization pass to divide the weighted sum by the sum of weights. Our work also presents a high-quality shader based on blending overlapping points, but without considering oriented disks with variable radii. The already discussed Auto Splats method~\cite{AutoSplats} uses the same high-quality rendering method, but computes the required attributes (normals and radii) on the fly.

\subsection{Level-of-Detail Rendering}

While our compute-based rendering method attempts to speed up the rendering of a given set of points, LOD methods aim to improve the performance by reducing the number of points being rendered. These goals are complementary and can be combined to further improve the overall point-cloud rendering performance. An LOD approach that is built on \enum{GL\_POINTS} may benefit from faster compute-based alternatives. LODs are addressed via sequential point trees~\cite{Dachsbacher2003}, layered point clouds~\cite{LPC} and their variations~\cite{Wand2008, Goswami2010, scheiblauer2011, Elseberg2013, MartinezRubi2015, Kang2019, Bormann:PCI}. We assess layered point cloud rendering with our methods in Section~\ref{sec:lod}.

\section{Compute-Shader Rendering}

In this chapter, we first describe a basic compute shader-based method for point cloud rendering, followed by various optimizations to improve performance and quality. A detailed evaluation of each variation is provided in Section~\ref{sec:evaluation}.

\begin{figure}
    \centering
    \hfill
    \begin{subfigure}[t]{\columnwidth}
        \includegraphics[width=\textwidth]{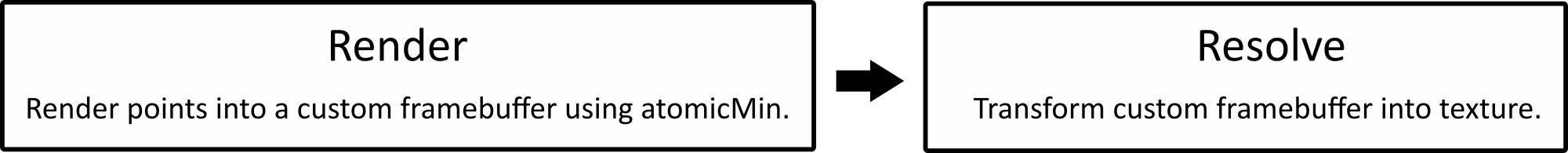}
        \caption{Our basic compute shader approach requires a render pass over all points, and a resolve pass over all pixels to transfer colors from the interleaved buffer into a texture or render target.}
    \end{subfigure}
    \hfill
    \begin{subfigure}[t]{\columnwidth}
        \includegraphics[width=\textwidth]{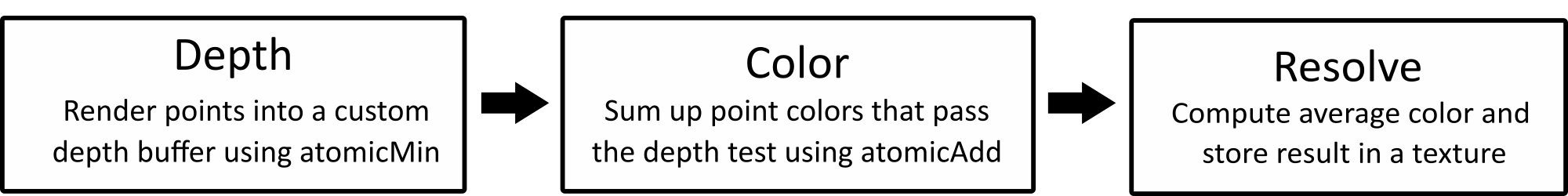}
        \caption{The high-quality approaches require two geometry passes (depth and color) over all points,  and a resolve pass over all pixels that writes blended color values into a displayable texture.}
    \end{subfigure}
    \caption{Proposed compute shader pipelines.}
    \label{fig:pipeline}
\end{figure}

Figure~\ref{fig:pipeline} shows the individual compute-shader passes for the basic and high-quality approaches. All proposed variations require at least one render pass over all points, and a resolve pass over all pixels. The render pass transforms points to screen space and writes them into a storage buffer that acts as a framebuffer. The resolve pass transfers the results from our custom framebuffer into an OpenGL texture. The high-quality variations further require an additional pass over all points to compute a depth map, for a total of two geometry passes and one screen pass. 

\subsection{Basic Approach}
\label{sec:atomicMin}

Our compute-based approach resolves visibility and color simultaneously by encoding the depth and color in a single 64bit integer, and uses 64bit \code{atomicMin} to pick points with the lowest depth, as shown in the following GLSL sample:

\begin{lstlisting}[language=Java,label=lst:basic,caption={Render pass draws closest points via \code{atomicMin}.},captionpos=b]
vec4 pos = worldViewProj * position;

int pixelID = toPixelID(pos);
int64_t depth = floatBitsToInt(pos.w);
int64_t point = (depth << 24) | rgb;

atomicMin(framebuffer[pixelID], point);
\end{lstlisting}

The compute shader replaces the role of both vertex and fragment shader in the standard rendering pipeline. In Listing \ref{lst:basic}, we only write the vertex color, stored in variable \textit{rgb}, to the framebuffer, as is common when rendering point clouds. Alternatively, we may also compute illumination if normals are present, or any other mapping from attributes to RGB color values. The target of the render pass is an OpenGL storage buffer containing one 64bit integer per pixel. For each point, the color value is stored in the 24 least significant bits of the integer (8 bits per color channel), and the 32bit depth value is stored in the 32 next-higher bits. Using \code{atomicMin}, we can then make sure that the point with the smallest depth value is chosen for each pixel. After the render pass, we transfer the results from this storage buffer into an OpenGL texture with a resolve pass over all pixels:

\begin{lstlisting}[language=Java,label={lst:resolve},caption={Resolving custom framebuffer to OpenGL texture.},captionpos=b]
uint pixelID = x + y * imageSize.x;
// Read RGB component from framebuffer
uvec4 icolor = colorAt(pixelID);
// Write result into an OpenGL texture
imageStore(texture, ivec2(x, y), icolor);
\end{lstlisting}

One beneficial side effect of this approach is the ability to directly use the linear depth value (\textit{pos.w}) for depth-buffering. In contrast, depth values in the conventional hardware rendering pipeline are subject to further transformations and associated loss of precision. In order to allow perspectively correct triangle rasterization, the typically used transformation matrices result in hyperbolic depth buffer values which, when combined with floating-point data types, causes a disproportionately larger accumulation of precision towards the near plane. We note that methods such as reverse-z~\cite{Reed18,OptimalDepthBuffer} can help achieve a near-optimal precision of the hardware pipeline by mapping the far plane to 0 and the near plane to 1. Nevertheless, since we don't require interpolation across pixels, we can directly use the unmodified linear depth value to obtain the maximum depth-buffer precision that a typical floating point transformation matrix may achieve. If traditional depth-buffer values are required (e.g., for hybrid rendering with triangles), developers may integrate the associated conversions in the resolve pass.

\subsection{Early Depth Test}
\label{sec:early-z}

\begin{figure}
    \centering
    \hfill
    \begin{subfigure}[t]{0.49\columnwidth}
        \includegraphics[width=\textwidth]{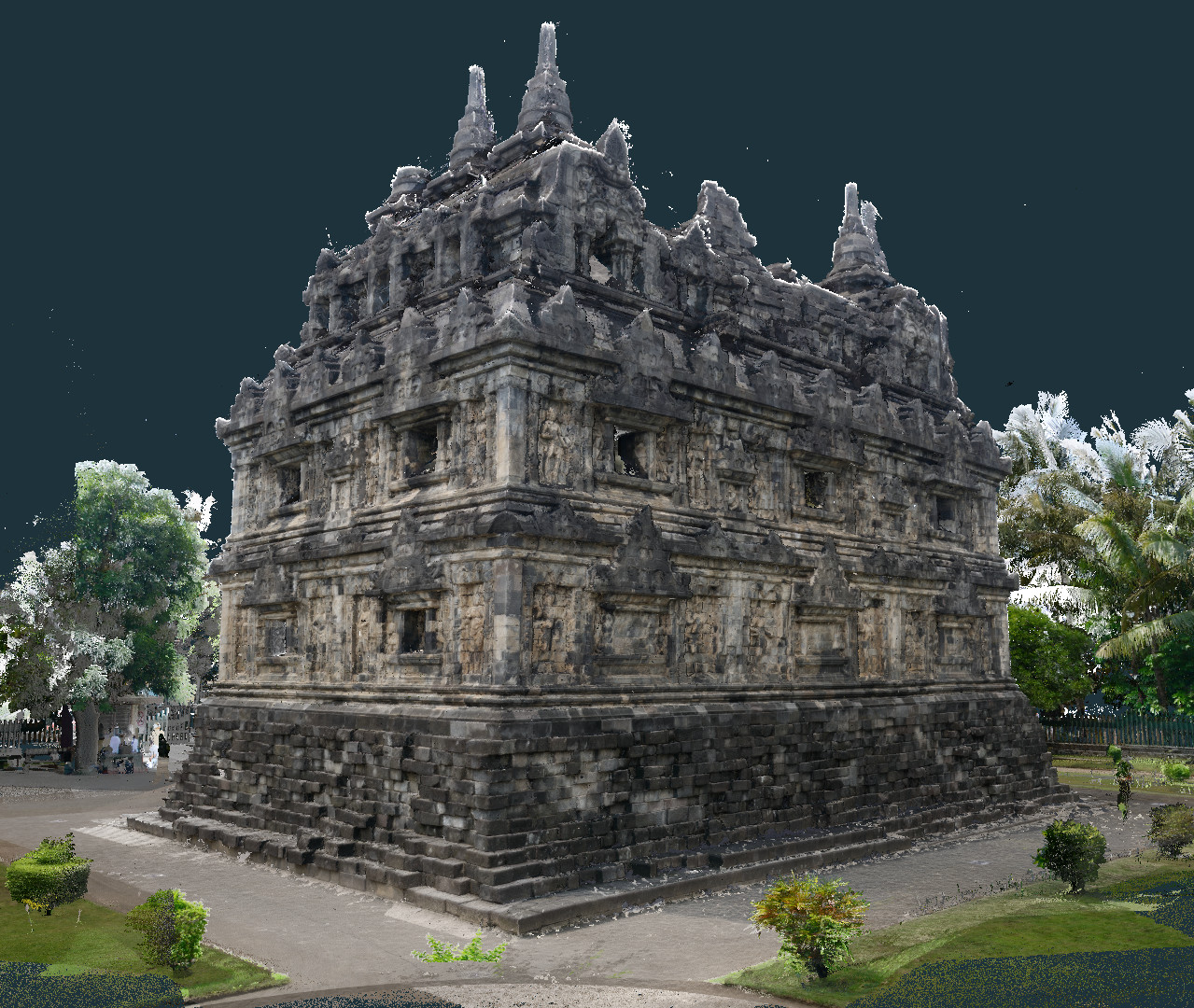}
        \caption{RGB}
    \end{subfigure}
    \hfill
    \begin{subfigure}[t]{0.49\columnwidth}
        \includegraphics[width=\textwidth]{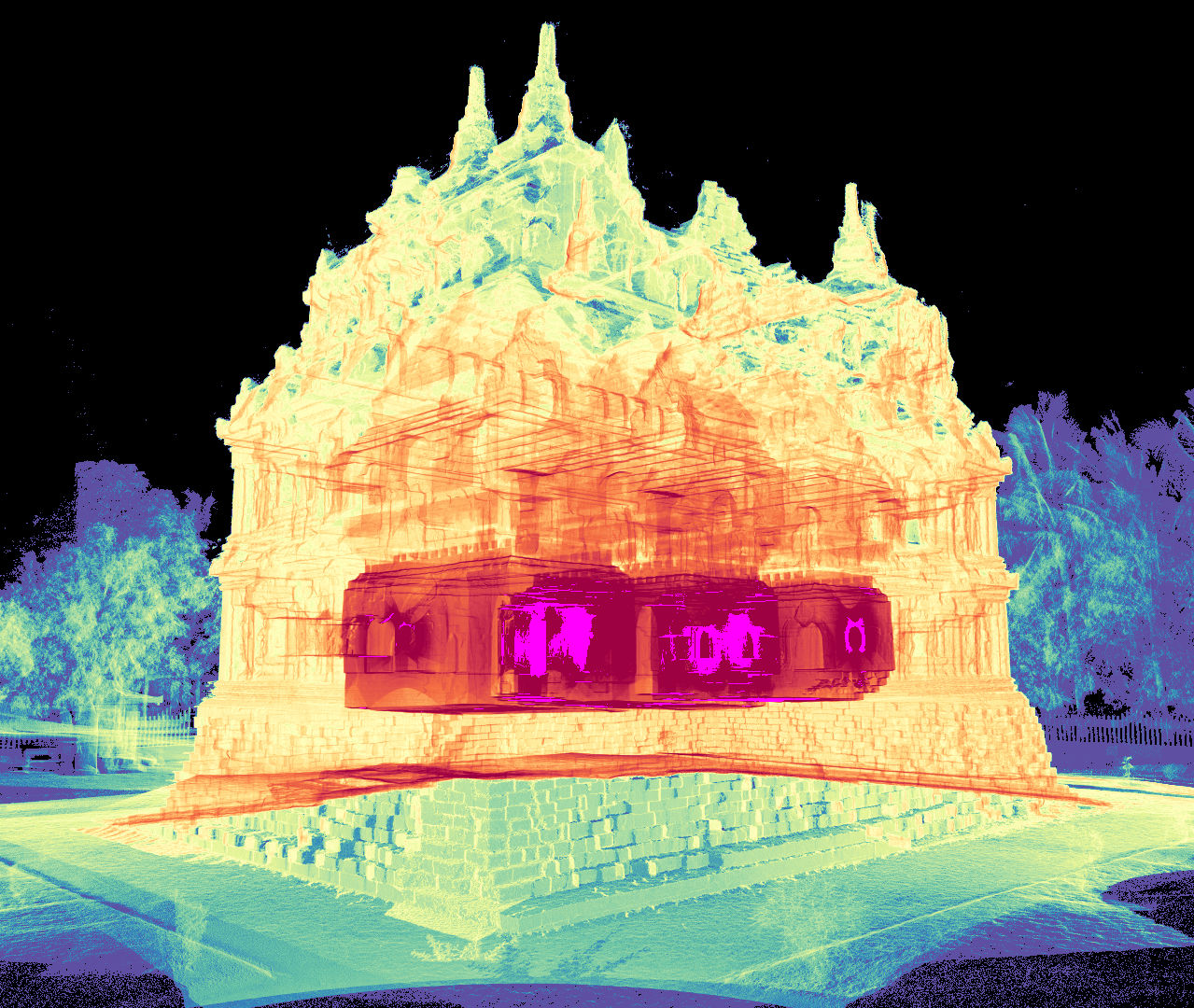}
        \caption{Points per pixel}
    \end{subfigure}
    \caption{(a) Candi Sari point cloud,  courtesy of TU Wien--Baugeschichte.  (b) Heatmap of the number of points inside each pixel. 9,427 pixels contain more than 10k points (highlighted in pink). Points: 725M. Resolution: 1920x1080. }
    \label{fig:fragcounts}
    \vspace{-0.3cm}
\end{figure}

Rendering hundreds of millions of points to framebuffers with approximately 2 million target pixels (1920x1080) can lead to the projection of several thousand points to a single pixel, depending on viewpoint and scan density distribution, as shown in Figure~\ref{fig:fragcounts}. Such a large number of atomic operations on the same memory location causes contention with severe negative impacts on the performance. In practice, this manifests as lower framerates after zooming out, even if the number of points within the view frustum stays the same. In the traditional rendering pipeline, early-z (also known as early depth test or early fragment test) is done before the fragment shader to avoid calling the shader program for fragments that are occluded by previously processed fragments. In our compute shader-based approach, we can achieve early depth testing by comparing against the current value at the storage buffer location we are trying to write to. \code{atomicMin} only needs to be called if the currently processed point has a smaller depth value than a previously written point. A GLSL snippet for early depth testing is shown below:

\begin{lstlisting}[language=Java,caption={Avoiding \code{atomicMin} calls with an early depth test.},captionpos=b]
uint64_t oldPoint = framebuffer[pixelID];

if(point < oldPoint)
	atomicMin(framebuffer[pixelID], point);
\end{lstlisting}

In contrast to~\cite{Gnther2013AGP}, our simplified solution exploits the fact that previously written depth values may already be available for reading in fast L1 caches.
Although loading and evaluating the current depth is not synchronized, this does not affect the rendered image because the real depth buffer value can only become smaller in the meantime.
In the worst case, we merely invoke superfluous \code{atomicMin} calls. 
%This policy can reduce the number of contended accesses to the slower L2 cache and device memory significantly. 

\subsection{Local Reduction with Warp-Level Primitives}
\label{sec:ballots}

Individual GPU threads are grouped in concurrently scheduled warps (NVIDIA), wavefronts (AMD) or subgroups (OpenGL). On NVIDIA microarchitectures, each warp consists of 32 threads that can operate based on the SIMT (single instruction, multiple threads) principle. Threads within each warp can communicate efficiently, opening up the possibility to combine intermediate results. In our case, we can combine up to 32 points into a single framebuffer update if they fall inside the same pixel, thereby reducing the amount of expensive \code{atomicMin} calls from 32 to just one.

In Listing~\ref{lst:ballots}, all threads in the subgroup first check if they have the same target pixel ID. If this is the case, the threads compute the minimum depth value of the whole subgroup and only the thread with the smallest depth proceeds to write a point to the framebuffer. The final \texttt{if} clause applies to the slow and the fast, reduction-based path, causing 32 updates in the former and a single update in the latter.

\begin{lstlisting}[language=Java,label={lst:ballots},caption={If all points in a warp fall into the same pixel, \code{atomicMin} is only called in the thread with the lowest depth. Otherwise, atomicMin must be called by all threads.},captionpos=b]
int minDepth = depth;

if(subgroupAllEqual(pixelID))
  minDepth = subgroupMin(depth);

// Different pixels or thread has lowest depth
if(minDepth == depth)
  atomicMin(framebuffer[pixelID], point);
\end{lstlisting}

The warp-wide reduction mostly affects very dense point clouds and zoomed-out views. In order to accelerate other cases as well, we can perform a quick, fine-granular local reduction where pairs of threads check whether their updates can be merged (see Listing~\ref{lst:latestage}). Doing so can help to reduce the number of atomic operations, even for close-up viewpoints, depending on the order of vertices in memory.

\begin{lstlisting}[language=Java,label={lst:latestage},caption={If two neighboring threads target the same pixel, only the one with the lower depth value will write its result. },captionpos=b]
uint idXOr = subgroupClusteredXor(pixelID, 2);
uint minDepth = subgroupClusteredMin(depth, 2);

// Update if different pixels or lowest depth
if(idXOr != 0 || minDepth == depth)
	atomicMin(framebuffer[pixelID], point);

\end{lstlisting}

\subsection{Full Warp-Wide Deduplication }
\label{sec:dedup}

Taking advantage of recent GLSL language extensions, we can fully deduplicate the pixel updates within a subgroup, i.e., we can further reduce framebuffer updates to a single \code{atomicMin} per accessed pixel, as shown in Listing~\ref{lst:dedup}.

\begin{lstlisting}[language=Java,label={lst:dedup},caption={The deduplication shader can reducs pixel updates to only one per accessed pixel in each SIMT subgroup.},captionpos=b]
// Find subgroup (sg) indices with same pixelID
uvec4 sg = subgroupPartitionNV(pixelID);

// Lowest depth among threads with same pixelID
int minDepth;
minDepth = subgroupPartitionedMinNV(depth, sg);
// Write if thread owns point with lowest depth
if(depth == minDepth)
	atomicMin(framebuffer[pixelID], point);
\end{lstlisting}

Each thread first requests a mask of all threads that intend to write to the same pixel. This mask is then used to compute the minimum depth of the respective threads, and only the threads with the lowest depth will update the pixel. Note that there may still be superfluous \code{atomicMin} calls if two or more points in a warp have the same pixel ID and depth value. It is possible to remove these duplicates, but in practice, we found no discernible benefits from doing so.

\subsection{High-Quality Rendering}
\label{sec:hqs}

While the presented approaches are fast and require only one pass over the entire geometry, they, like all single-pixel update methods, exhibit noticeable aliasing. Point-cloud renderings are colored per-vertex, which causes artifacts similar to rendering textured models without mipmapping. Mipmapping addresses the cases where the bounds of a single pixel map to a large area of a texture. Instead of picking a single texture element (texel), all texels contained in a pixel should contribute to that pixel. Mipmaps store precomputed averages of potentially contributing texels to avoid recomputing averages of potentially thousands of texels in each frame. Unfortunately, mipmapping is not feasible in our case: two-dimensional textures are not applicable to regular point-cloud data sets due to the lack of 2D neighborhood information. Instead, we propose a compute shader-based high-quality approach to calculate the average of overlapping points inside a pixel directly at runtime. 

\begin{figure}
    \centering
    \hfill
    \begin{subfigure}[t]{0.49\columnwidth}
        \includegraphics[width=\textwidth]{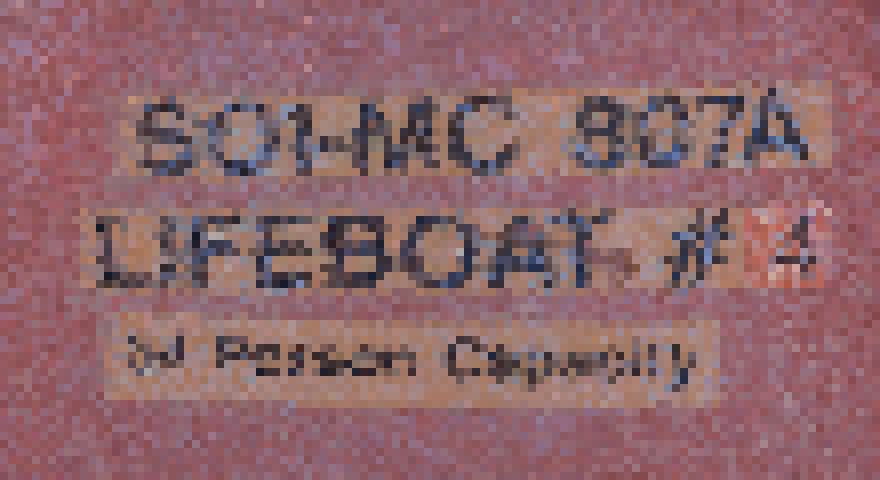}
        \caption{Basic Shaders}
    \end{subfigure}
    \hfill
    \begin{subfigure}[t]{0.49\columnwidth}
        \includegraphics[width=\textwidth]{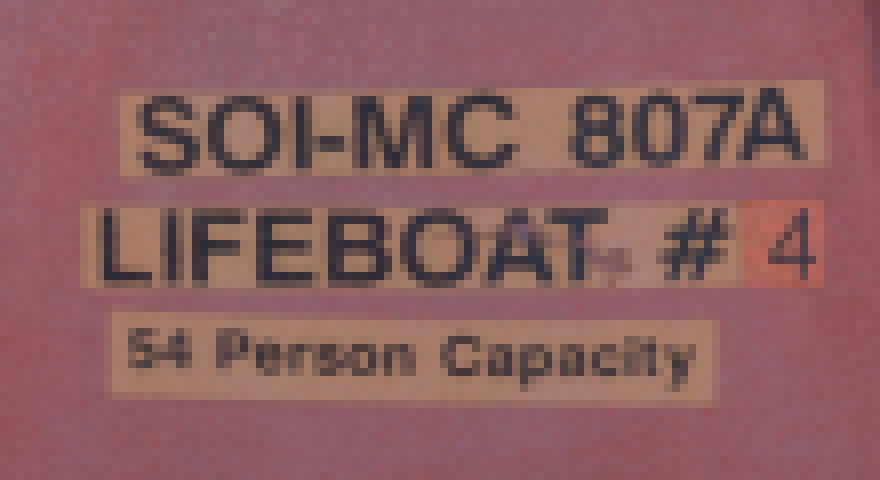}
        \caption{HQS Shaders}
    \end{subfigure}
    \caption{(a) \enum{GL\_POINTS} and our basic \code{atomicMin}-based approaches are prone to aliasing artifacts. (b) High-quality shading blends overlapping points together, improving quality and legibility of high-frequency features, such as text. Lifeboat point cloud courtesy of Weiss AG.}
    \label{fig:aliasing}
\end{figure}

One concept for high-quality shading is that all overlapping points of the front-most visible surface should contribute to the pixel color, rather than just the point that is closest to the camera. Botsch et al.~\cite{Botsch:HQS} achieve this through a multi-pass approach that first computes a depth map, which is then shifted slightly towards the far plane. They then sum up the colors of fragments that pass the depth test, and finally obtain the average by dividing the sum of all colors by the number of fragments that contributed to the sum. Figure~\ref{fig:aliasing} illustrates the difference between basic approaches, including \enum{GL\_POINTS}, and such a high-quality shading (HQS) approach. HQS reduces image noise by computing a blend of all relevant samples, and increases the fidelity of high-frequency features that might otherwise be lost.
%because they were not among the single points in each pixel that was accepted by one of the basic approaches.
In order to apply this method on the GPU, we employ two geometry render passes---a depth and a color pass---over all points. The depth pass is implemented similar to the previously introduced \code{atomicMin} approaches, but without the need to update colors. We therefore implement it with simple 32bit atomic operations instead of 64bit. 

In the color pass, we replace \code{atomicMin} by \code{atomicAdd} in order to sum up the values of all contributing points and their count. Contributing points should be those that belong to the front-most visible surface. However, since there is no notion of a surface in a point cloud, we instead propose considering all points within an $\epsilon$ range, proportional to the distance of the closest point. In practice, we obtained good results with an $\epsilon$ of 1\%. To provide sufficient capacity for computing the sum of a large amount of points, we recommend a buffer with 32 bits per color channel, and a fourth channel for the fragment count. Summing up all channels requires four 32bit \code{atomicAdd} calls, but we suggest packing the channels into two 64bit integers to reduce the number of \code{atomicAdd} updates to two per point, as shown in Listing \ref{lst:hqs_render}. The resolve pass then divides the color sums by the total number of points that contributed to them (Listing \ref{lst:hqs_resolve}).
\begin{lstlisting}[language=Java,label={lst:hqs_render},caption={Accumulate colors and point count within an  $\epsilon$ distance of the closest point (1\% farther away, in this case).},captionpos=b]
uint bufferVal = ssDepthbuffer[pixelID];
float bufferDepth = uintBitsToFloat(bufferVal);
float depth = pos.w;

if(depth <= bufferDepth * 1.01){
	int64_t rg = (r << 32) | g;
	int64_t ba = (b << 32) | 1;

	atomicAdd(ssRG[pixelID], rg);
	atomicAdd(ssBA[pixelID], ba);
}
\end{lstlisting}
\begin{lstlisting}[language=Java,label={lst:hqs_resolve},caption={Normalize to the average color in the resolve pass.},captionpos=b]
uint64_t rg = ssRG[pixelID];
uint64_t ba = ssBA[pixelID];

uint a = uint(ba & 0xFFFFFFFFUL);
uint r = uint((rg >> 32) / a);
uint g = uint((rg & 0xFFFFFFFFUL) / a);
uint b = uint((ba >> 32) / a);
\end{lstlisting}

\subsubsection{Single-Atomic High-Quality Variants}
\label{sec:hqs_1xatomic}

The number of atomic operations in the high-quality shader can be further reduced by encoding the RGB channels and counter into a 64bit integer. A distribution of 18 bits for the RGB channels and 10 bits for the counter guarantees that we can correctly compute the sum of up to 1023 points per pixel with one 64bit atomic operation per point (HQS1$\times$).

\begin{figure}
    \centering
    \hfill
    \begin{subfigure}[t]{0.49\columnwidth}
        \includegraphics[width=\textwidth]{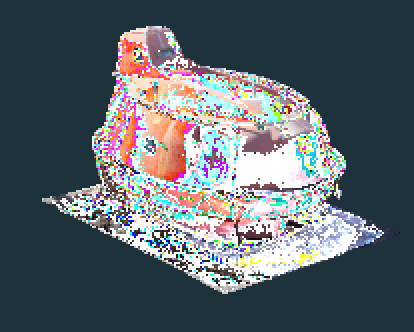}
        \caption{HQS ($1\times$atomicAdd)}
    \end{subfigure}
    \hfill
    \begin{subfigure}[t]{0.49\columnwidth}
        \includegraphics[width=\textwidth]{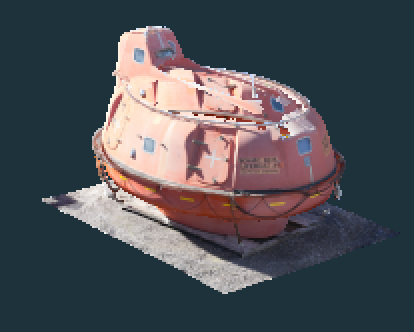}
        \caption{HQS (robust $1\times$atomicAdd)}
    \end{subfigure}
    \caption{(a) The $1\times$\code{atomicAdd} shader is prone to artifacts when a pixel receives more than 1023 points. (b) The robust version adds an overflow protection that eliminates these artifacts. Lifeboat point cloud courtesy of Weiss AG.}
    \label{fig:hqs_artifacts}
\end{figure}

While this na\"{i}ve approach may work well for small point clouds and close-up views, the counter and color values are subject to overflow whenever more than 1023 points contribute to a pixel's average color value. For larger scenes, this leads to strongly noticeable rendering artifacts, as shown in Figure~\ref{fig:hqs_artifacts}. Hence, we propose a robust alternative: instead of a 18-18-18-10 fragmentation, we use a 16-16-16-16 pattern. This ensures that up to 255 8bit color values can be stored in RGB, and the counter can register $2^{16} = 65\,536$ increments. 
We further modify the shader to check the return value after performing the 64bit atomic addition. If the count exceeds 255, the update is written a second time into separate fallback buffers with two atomic operations, as described in Section \ref{sec:hqs}.
The thread that detected an exact count of 255 is further tasked with transferring the pixel data to the fallback buffers and atomically setting the full 64bit value to zero.
The unused $2^{16}-256$ counts merely serve as a safety margin for---potentially several thousand---simultaneously executing threads to detect the overflow. If more or significantly fewer than 65k simultaneous accesses to a pixel are expected, the bit pattern may be changed accordingly. In practice, however, we found that this robust variant, \textit{HQS1R}, with a $4\times16$ pattern shows negligible overhead compared to the unstable HQS1$\times$ version.
Listing \ref{lst:hqs_1x64bit_robust} shows the method's implementation.
The associated resolve pass must then combine the results from all updated buffers.

\begin{lstlisting}[language=Java,label={lst:hqs_1x64bit_robust},caption={Summing up colors with a 16-16-16-16 bit pattern per point for the r-g-b-a channels and fallback on overflow. },captionpos=b]
uint64_t rgba = (r<<48)|(g<<32)|(b<<16)|1;
// Try rendering with just one atomicAdd
uint64_t old = atomicAdd(ssRGBA[pixelID],rgba);
uint count = uint(old & 0xFFFF);

if (count >= 255){
	// Overflow detected
	int a = 1;
	if (count == 255){
		// The first thread that overflows resets 
		// the overflown buffer value to zero, and
		// transfers the r, g, b, a values before
		// the overflow to the fallback buffer.
		atomicExchange(ssRGBA[pixelID], 0);
		r += (old>>48) & 0xFFFF;
		g += (old>>32) & 0xFFFF;
		b += (old>>16) & 0xFFFF;
		a += 255;
	}
	int64_t rg = (r << 32) | g;
	int64_t ba = (b << 32) | a;
	// All overflown threads write the value 
	// to the fallback buffer using 2 atomicAdd
	atomicAdd(fallback[2 * pixelID + 0], rg);
	atomicAdd(fallback[2 * pixelID + 1], ba);
}
\end{lstlisting}

\section{Vertex Order Optimization}

Beyond various compute-based solutions for point rendering, we also investigate the impact of point order in memory. Reordering vertices and indices is a common task for mesh optimization prior to rendering and can significantly improve data access patterns. Similarly, point clouds can be stored in various different vertex orders. Since we use non-indexed draws, clearly the fetching of point-cloud data is already optimally coalesced. However, we can also attempt to achieve a point ordering that positively impacts the updates of the framebuffer itself. One potential point ordering method is to shuffle them randomly. Shuffled or partially shuffled data sets can be used for progressive rendering~\cite{schuetz-2020-PPC} or as a simple level-of-detail structure, where rendering increasingly larger random subsets is akin to rendering increasingly higher levels of detail. Unfortunately, shuffling and the consequent absence of all data locality can be detrimental to rendering performance: Distributed writes to random framebuffer locations quickly lead to cache thrashing and fail to exploit the GPU's available memory bandwidth. A common method for increasing locality without the use of spatial data structures is to sort by Morton code. In the resulting Z-curve, points that are stored side-by-side in memory are also likely to be close in 3D and, assumedly, also in 2D after projection, thus favoring aggregate updates of close-by framebuffer locations. 

While ordering points by Morton code may in fact be adequate for compute variants that write to buffers backed by linear memory, it is suboptimal for the hardware rendering pipeline: for parallel rasterization, all output geometry is sorted into 2D viewport tiles. On NVIDIA architectures, each tile is statically assigned to a graphics processing cluster (GPC) with exclusive access to the underlying memory. GPC tiles are spread over the viewport in repeating 2D patterns to facilitate uniform load balance. However, it has been shown that these static patterns cannot compensate for extreme spatial clustering of in-flight scene geometry~\cite{3105762.3105777}. 
%Thus, neither shuffling nor Morton code yields an optimal ordering for the hardware pipeline.
To simultaneously address the hardware pipeline's demands for locality and sufficient utilization of all available GPCs, we thus propose an alternative layout that requires minimal overhead and is easy to produce: shuffled Morton order. 
First, all points are sorted according to Morton code. The resulting sequence is divided into batches of spatially sorted points. Batches are then moved to random locations without changing their internal order.
The final sequence yields uniformly distributed groups of spatially close points. 
This new ordering accurately recreates the typical workload of the rasterization engine, i.e., geometry primitives that provoke an update for a localized set of fragments.
In practice, we found that shuffled Morton order drastically improves performance when rendering with \enum{GL\_POINTS}. 
As an unexpected side effect, it also tends to reduce fluctuations and performance gaps across different compute-based methods (see following Section \ref{sec:evaluation} for details).
%For details, please see Section \ref{sec:evaluation}, where we assess each scene in our test suite using its original, shuffled, Morton and shuffled-Morton order. 

\section{Evaluation}
\label{sec:evaluation}

In this section, we evaluate our proposed compute shader-based point cloud rendering approaches and compare them against each other, as well as adequate baselines under relevant aspects that can affect the performance of rendering large point clouds. 

The evaluated methods are:
\begin{itemize}
    \item \textbf{GL\_POINTS}: Hardware pipeline using OpenGL \enum{GL\_POINTS} primitive. Points are rendered with the default size of one pixel. The \textit{gl\_PointSize} variable is unmodified.
    \item \textbf{just-set}: A reference method with no program control overhead or performance penalties for memory synchronization. A compute shader draws points by setting the framebuffer's pixel colors in a non-atomic fashion. Cannot be used in practice, due to the lack of depth buffering and resulting flickering due to the unpredictable order in which points are written by multiple compute threads.
    \item \textbf{busy-loop}: Implementation of the busy-loop approach by Günther et al.~\cite{Gnther2013AGP}, including their early-z test.
    \item \textbf{atomicMin}: Our basic compute-shader approach, which uses \code{atomicMin} to store the point with the smallest depth inside the framebuffer, as described in Section~\ref{sec:atomicMin}.
    \item \textbf{early-z}: The basic \textbf{atomicMin} method, but with an additional early-z test, as described in Section~\ref{sec:early-z}.
    \item \textbf{reduce}: The basic \textbf{atomicMin} method, but with an additional warp-wide reduction to decrease the number \code{atomicMin} calls from 32 to 1 if all points in the warp are inside the same pixel, or by up to 50\% if pairs of adjacent threads write to the same pixel (outlined in Section~\ref{sec:ballots}). 
    \item \textbf{reduce\&early-z}: Applies both, early-z and reduce.
    \item \textbf{dedup}: A full deduplication shader, that reduces the amount of \code{atomicMin} calls in a subgroup to one per pixel (except if depth values are identical). See Section~\ref{sec:dedup} for details. This method also includes an early-z test. 
    \item \textbf{HQS}: \textit{High-Quality Shading}. Computes average color of overlapping points in a certain depth range (Section~\ref{sec:hqs}).
    \item \textbf{HQS1R}: Robust high-quality shading with overflow protection using one \code{atomicAdd} in its fast path, and two atomic updates as fallback, as described in Section~\ref{sec:hqs_1xatomic}. 
\end{itemize}

Furthermore, the order in which points are stored in the vertex buffer has a significant impact on rendering performance---up to an order of magnitude between differently sorted point clouds. The evaluated orderings are:
\begin{itemize}
    \item \textbf{Original}: Points are kept in the order in which we originally received them. Typical orderings vary from sorted by Morton code, by scan position and timestamp, or even partially shuffled to simulate coarse LODs. 
    \item \textbf{Morton}: The point cloud is sorted by a 21 bit Morton code~\cite{Morton}---equivalent to sorting points using depth-first traversal into an octree with a depth of 21 levels. 
    \item \textbf{Shuffled}: Points are shuffled by assigning a random value to each point, and then sorting according to that value.
    \item \textbf{Shuffled Morton}: Points are first sorted by Morton code, then grouped into batches of 128 points, and finally the batches are shuffled, with points inside each batch remaining in order. Doing so preserves basic locality between points within each batch, but avoids excessive locality that might lead to contention or imbalanced workload.
\end{itemize}

Performance for each method and ordering is measured using OpenGL timestamp queries at the start and end of each frame (right before \code{glfwSwapBuffers}). Reported performance numbers are computed as the average of all frames over one second. In the following, we compare our methods and relevant baselines with respect to the most significant aspects, based on our observations on three test systems:
\begin{itemize}
    \item NVIDIA RTX 3090 24GB, AMD Ryzen 7 2700X (8 cores), 32GB RAM, running Microsoft Windows 10.
    \item NVIDIA RTX 2070 Super 8GB, Intel i7-4771 (8 cores), 16GB RAM, running Microsoft Windows 10.
    \item NVIDIA GTX 1060 3GB, AMD Ryzen 5 1600X (6 cores), 32GB RAM, running Microsoft Windows 10.
    %\item \textbf{940}: Intel Core i7-7500U CPU; NVIDIA 940MX 2GB; 16GB RAM.
\end{itemize}

However, due to the number of variables (method, vertex order, GPU), we will mainly focus on the illustration of results from the RTX 3090. Detailed benchmarks for the GTX 1060 and the RTX 2070 can be found in the paper appendix.

\begin{table*}[]
\begin{tabular*}{\textwidth}{ll}
\toprule
 \textbf{Model:} Lion & \textbf{Points:} 4 million (64MB) \\ 
 %\midrule
 \includegraphics[width=0.45\textwidth]{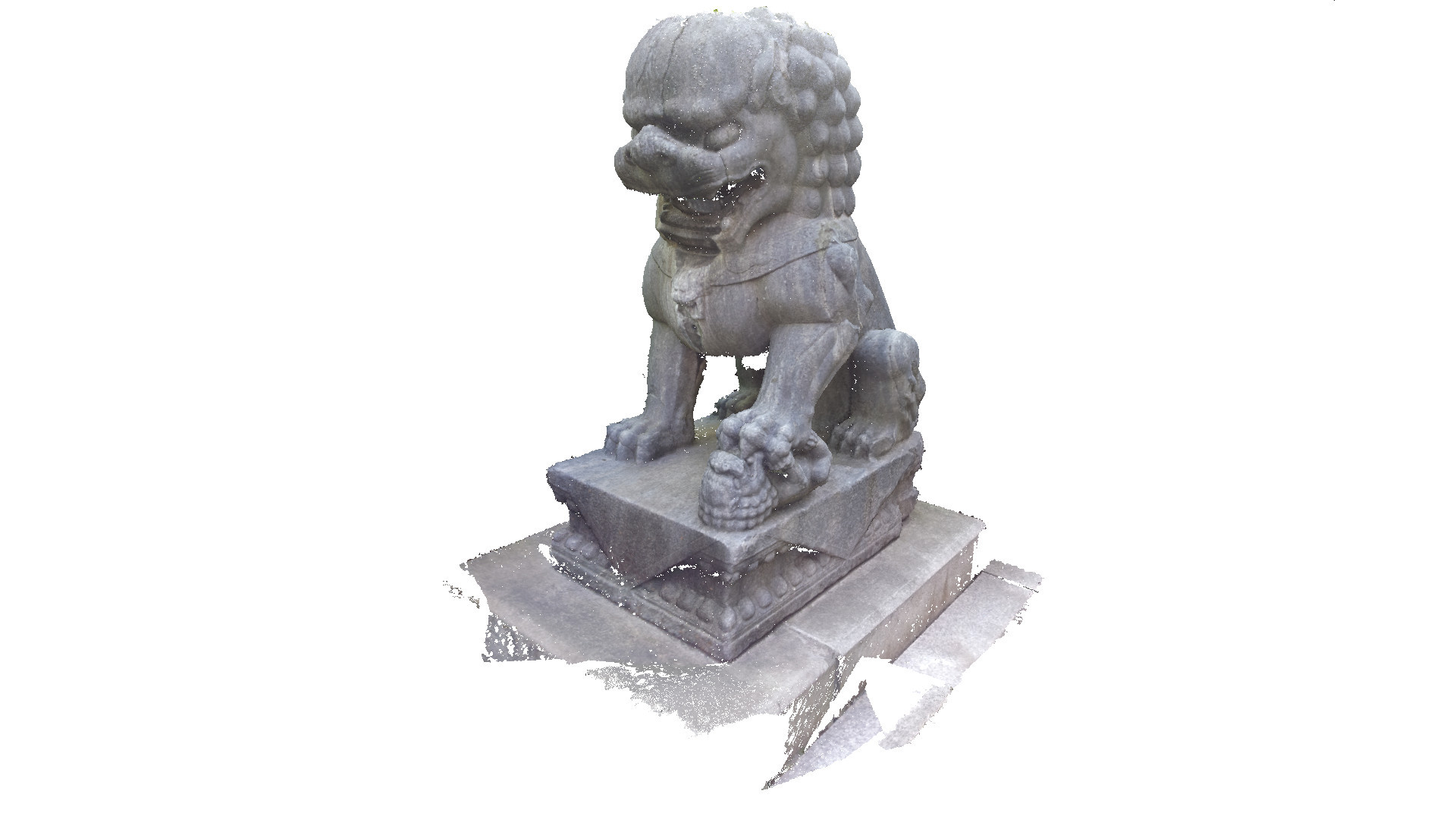} & 
 \includegraphics[width=0.5\textwidth]{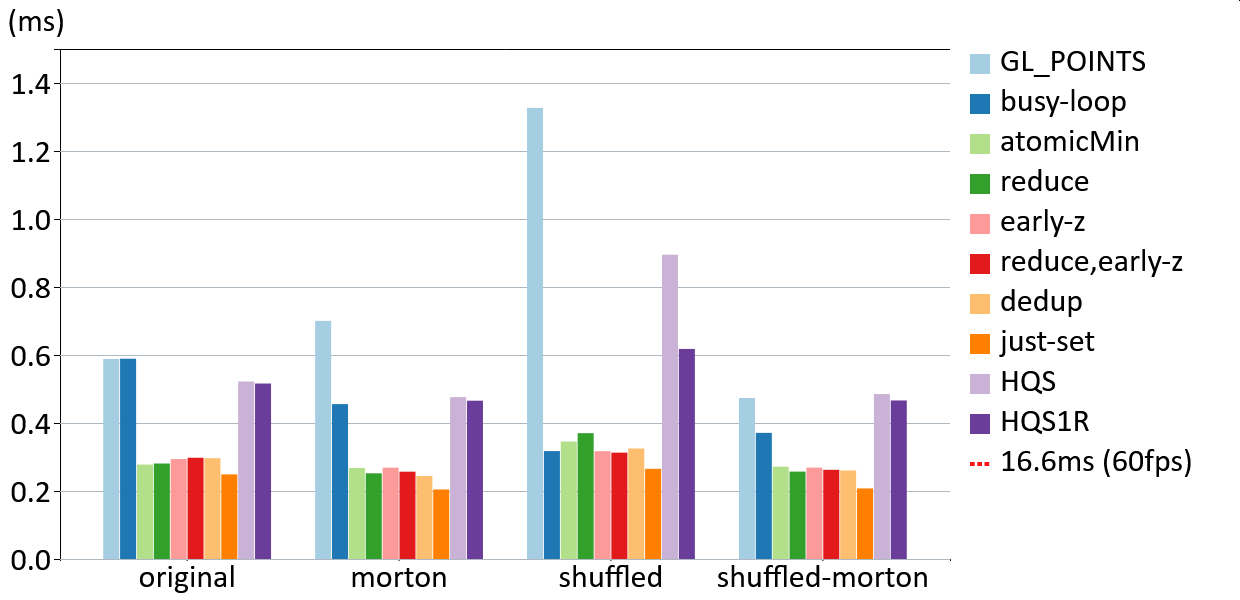} \\
 \midrule
 \textbf{Model:} Lifeboat & \textbf{Points:} 47 million (752MB) \\ 
 \includegraphics[width=0.45\textwidth]{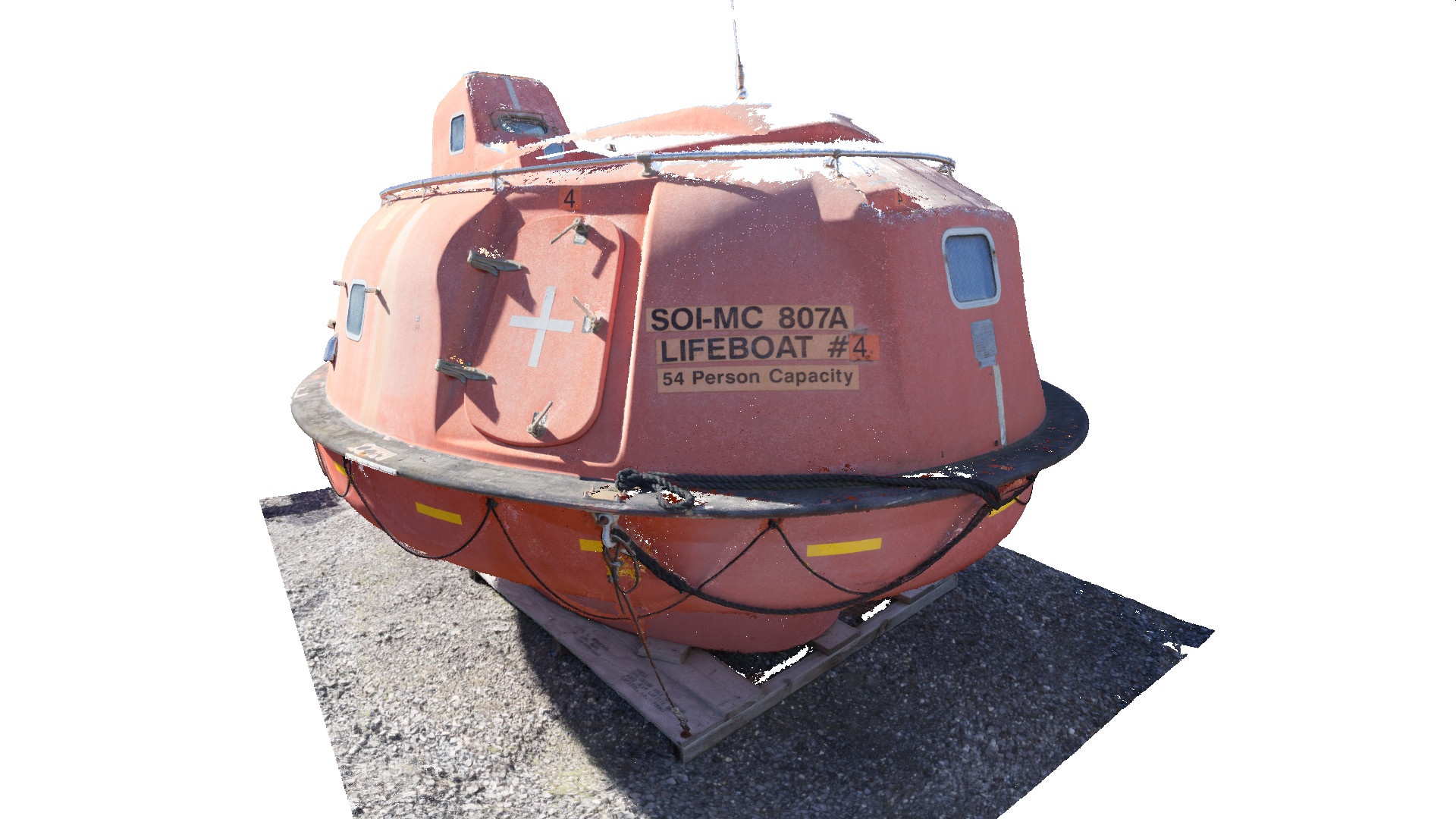} & 
 \includegraphics[width=0.5\textwidth]{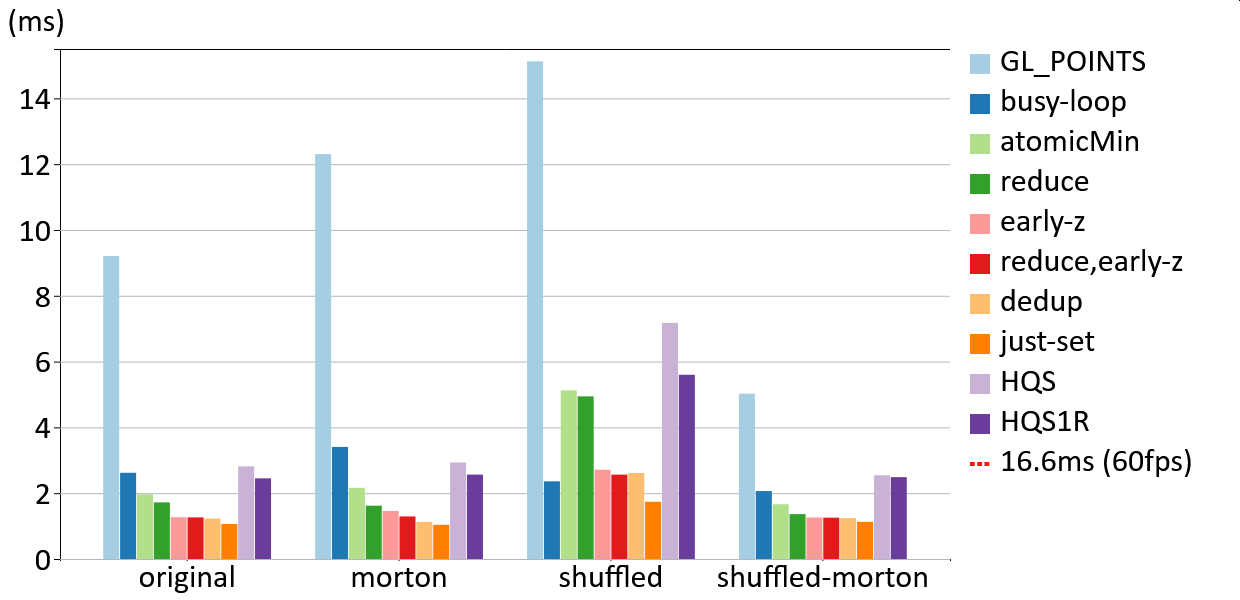} \\
 \midrule
 \textbf{Model:} Retz & \textbf{Points:} 145 million (2.3GB)  \\ 
 \includegraphics[width=0.45\textwidth]{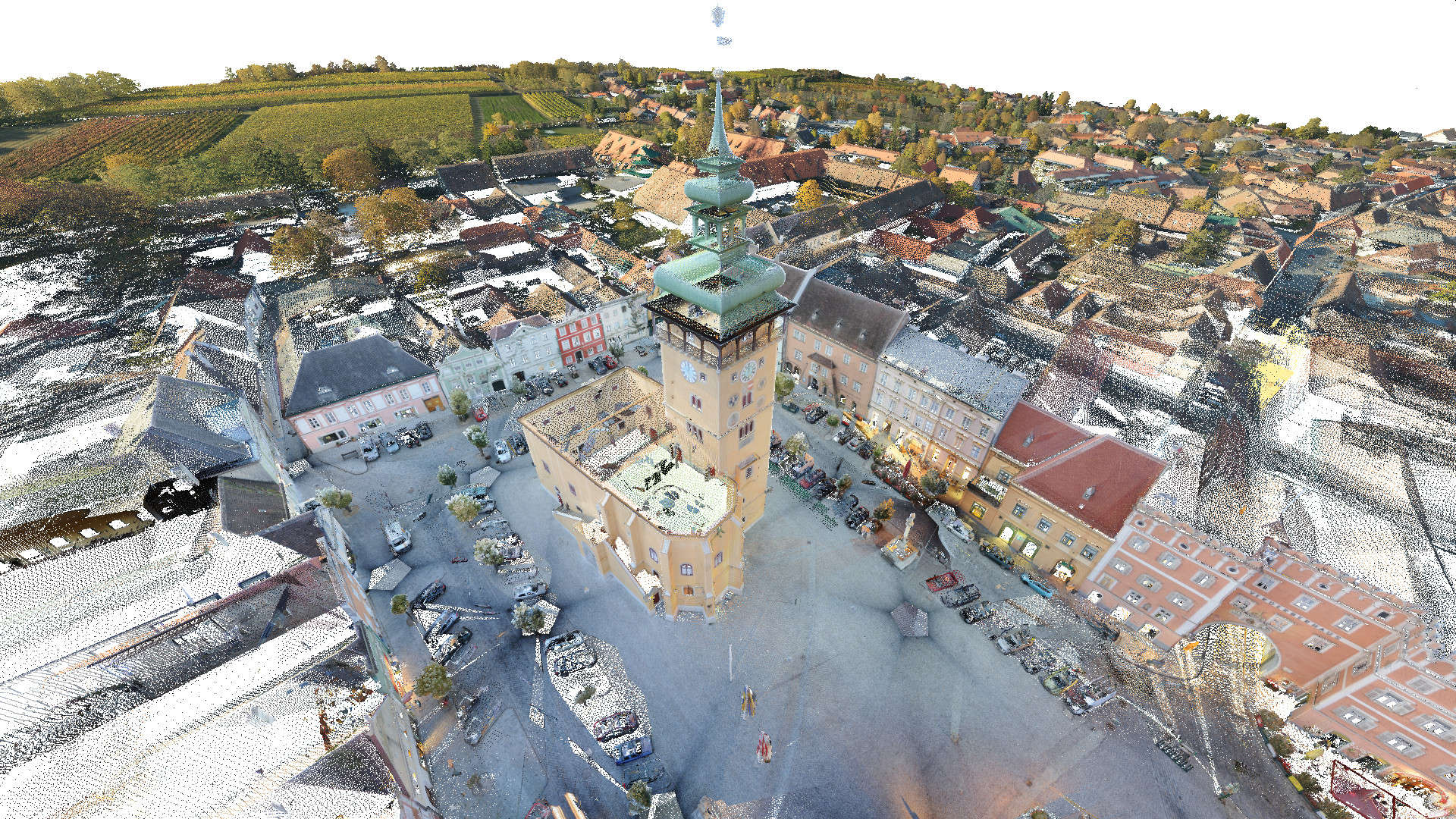} & 
 \includegraphics[width=0.5\textwidth]{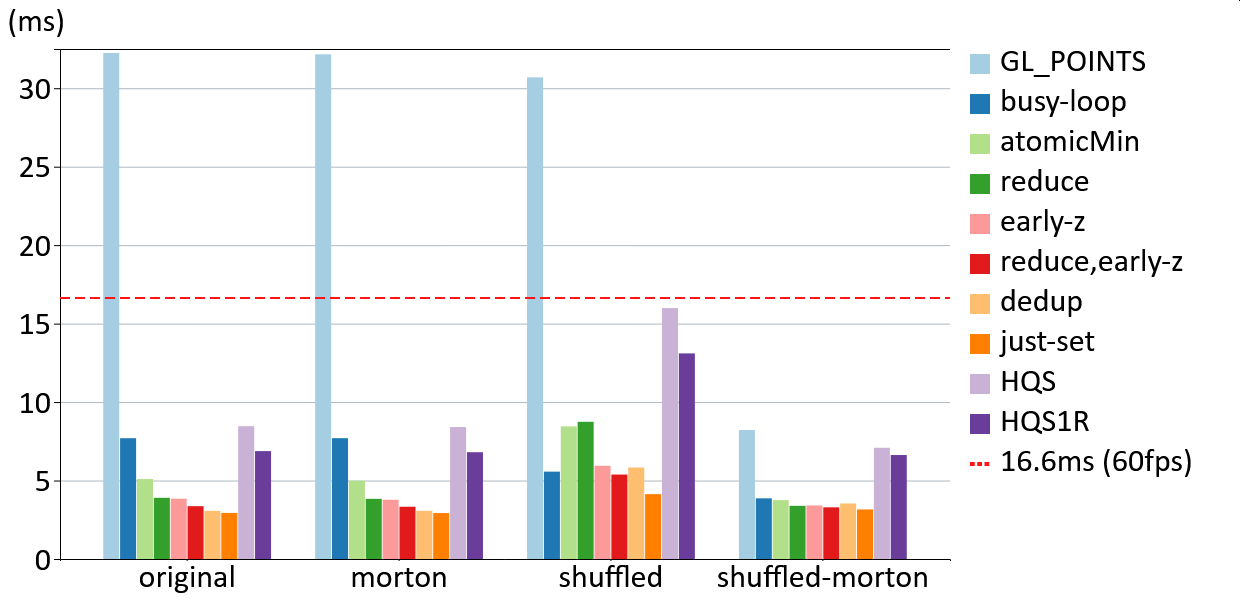} \\
 \midrule
 \textbf{Model:} Endeavor & \textbf{Points:} 796 million (12.7GB) \\ 
 \includegraphics[width=0.45\textwidth]{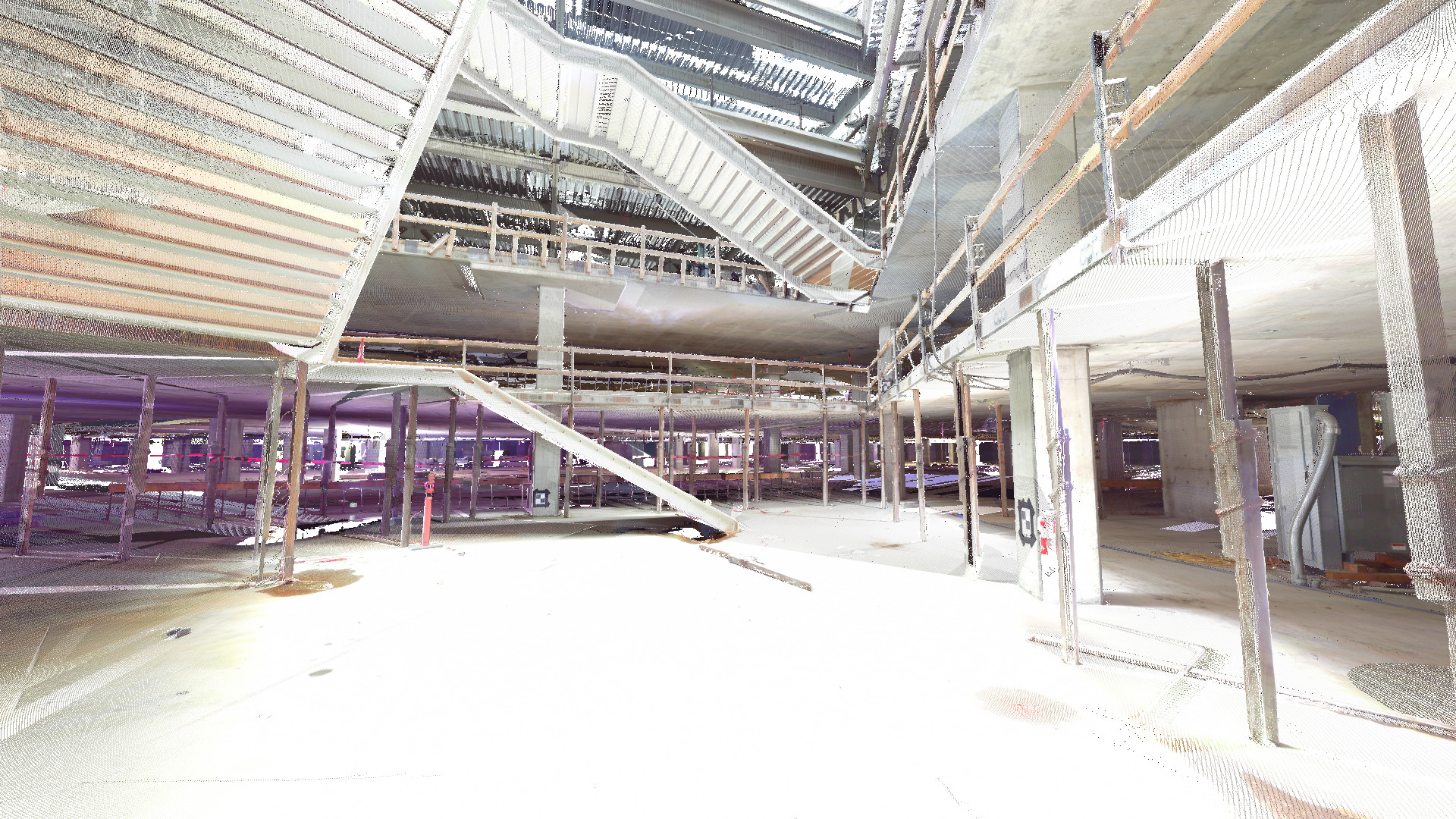} & 
 \includegraphics[width=0.5\textwidth]{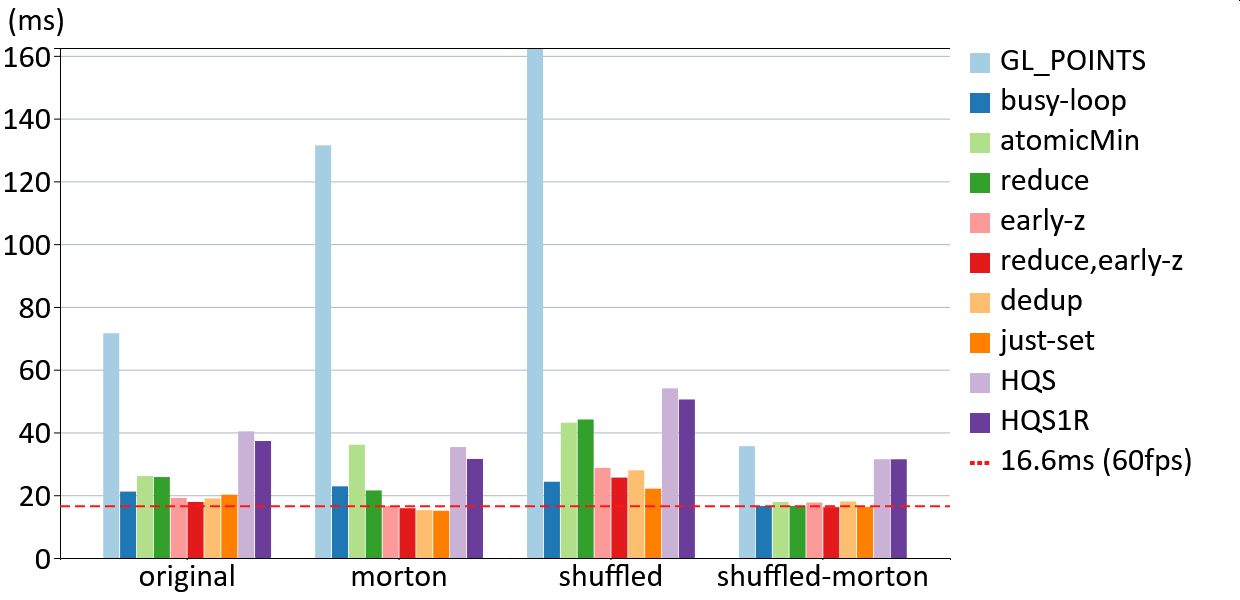} \\
\bottomrule
\end{tabular*}
\caption{Frame times (ms) for a given viewpoint and vertex order on an RTX 3090 (lower is better).}
\label{tab:scenarios}
\end{table*}

% \begin{table*}[]
% \begin{tabular*}{\textwidth}{ll}
% \toprule
%  \textbf{Model:} Retz (all points outside frustum) & \textbf{Points:} 145 million (2.3GB) \\ 
%  %\midrule
%  \includegraphics[width=0.45\textwidth]{images/retz_nopoints.jpg} & 
%  \includegraphics[width=0.5\textwidth]{images/charts/perf_retz_nopoints.png} \\
%  \midrule
%  \textbf{Model:} Retz (closeup) & \textbf{Points:} 145 million (2.3GB) \\ 
%  \includegraphics[width=0.45\textwidth]{images/retz_closeup.jpg} & 
%  \includegraphics[width=0.5\textwidth]{images/charts/perf_retz_closeup.png} \\
%  \midrule
%  \textbf{Model:} Retz (overview) & \textbf{Points:} 145 million (2.3GB)  \\ 
%  \includegraphics[width=0.45\textwidth]{images/retz_zoomed_out.jpg} & 
%  \includegraphics[width=0.5\textwidth]{images/charts/perf_retz_overview.png} \\
% \bottomrule
% \end{tabular*}
% \caption{Frame times (ms) for Retz in three different viewpoints on an RTX 3090 (lower is better). \enum{GL\_POINTS} and atomicMin are strongly affected by the viewpoint, while the combination of reduce and early-z is highly robust to changes of the viewpoint.}
% \label{tab:scenarios_viewpoint}
% \end{table*}

\subsection{Hardware vs. Compute Pipelines}

%\enum{GL\_POINTS} is the de-facto standard rendering primitive for high-performance rendering of large point clouds in practice.
Table~\ref{tab:scenarios} shows benchmark results comparing all evaluated methods for different data sets ranging from 4 million to 796 million points. The results clearly show that compute-based methods (including high-quality variants with additional anti-alisaing) are significantly faster than \textbf{GL\_POINTS}, the predominantly used method in state-of-the-art solutions, by a factor of 10$\times$ or more. The improvements are least pronounced for the lion model due to the implied overhead caused by the additional resolve pass, which takes about 0.06ms on an RTX 3090, and compute pipeline launches in general. %. Resolve pass invocations scale with the number of pixels rather than the number of points, and therefore their overhead becomes more notable as the number of pixels approaches the same order of magnitude as the number of points. 
After investigating with Nsight Systems, we found that the \textit{glDispatchCompute} call itself appears to have a larger overhead than \textit{glDrawArrays}, which is significant for small data sets, but negligible when rendering larger scenes. 

Among the individual compute-based methods, the \textbf{busy-loop} approach is often already outperformed by our most basic \textbf{atomicMin} and \textbf{reduce} methods, with the exception of particularly large or shuffled data sets. \textbf{early-z} can yield significant performance improvements (34\% on average), but has adverse effects on rendering the smaller lion model. This is expected, because early-z testing only pays off once the depth buffer is partially filled by previously processed points and thus mostly affects larger scenes. 
Synergizing the properties of \textbf{reduce} and \textbf{early-z} in \textbf{reduce\&early-z} yields a particularly strong and stable contender for all scenarios (1\% faster than fastest of either \textbf{reduce} and \textbf{early-z}, 33\% faster than slowest on average).
It is outperformed only on the RTX 3090 by the \textbf{dedup} method, whose frame time is 1--2\% faster on average and less affected by different vertex orderings.
%Our basic \textbf{atomicMin} approach without early-z already proves to be faster than the \textbf{busy-loop} approach (with early-z) in many cases, except for the Endeavor data set---likely because it contains so many points that early-z is required to compete. 
%Another important observation is that the reduce and the early-z methods each show improvements on their own, but the combination of both is typically as fast as the better one or faster.
In contrast to the \textbf{busy-loop} approach, which thrives on shuffled vertex order, the performance of our compute-based methods is best when using sorted or partially sorted data sets.
Specifically, the highest performance for every single scene was recorded on the RTX 3090, using \textbf{dedup} with Morton sorted scenes (at least 3\% faster than the next better technique).
\textbf{dedup} is most effective on the RTX 3090 and 2070, but clearly trails behind \textbf{reduce\&early-z} on the GTX 1060. This comes as no surprise, since the \code{subgroupPartitionNV} and \code{subgroupPartitionedMinNV} primitives in GLSL are only accelerated by GPU hardware since the Volta and Ampere architectures, respectively. In contrast, \textbf{reduce\&early-z} only requires warp shuffle instructions, which have been accelerated in hardware since the Kepler generation. 
For our high-quality variants, we found that \textbf{HQS1R} always runs faster than \textbf{HQS} (up to 37\%, 18\% on average). Both methods are up to 4$\times$ faster than the aliased \textbf{GL\_POINTS} (see appendix for visual comparison).

\subsection{The Impact of the Viewpoint on Performance}
\label{sec:viewpoint}

A key issue when rendering large point clouds is the spatial clustering of fragments, which can lead to contention and poor occupancy. The viewpoint plays an important role in that aspect. Close-up viewpoints of a model are typically advantageous, because many points fall outside the view frustum, but also because the points themselves are distributed more evenly, so the chance of individual pixels containing a large number of points at once is low. However, when zooming out, points start to cluster inside an exceedingly smaller range of pixels, some of which may now receive tens of thousands of points, as shown in Figure~\ref{fig:fragcounts}. We observe that the performance of rendering vertices in their original or Morton order with \textbf{GL\_POINTS} is strongly affected by the viewpoint.
Rendering the \textit{Retz} scene with a zoomed-in and zoomed-out view yields 31.95ms vs. 153.22ms ($\times4.79$), respectively.
The simple \textbf{busy-loop} and \textbf{atomicMin} approaches show similar trends (7.76ms vs. 16.66ms $\rightarrow \times2.15$ and 5.07ms vs. 21.45ms $\rightarrow \times4.23$), while sophisticated compute shader variants like \textbf{reduce\&early-z} and \textbf{HQS1R} are mostly unaffected (3.37ms vs. 3.33ms and 6.91ms vs. 7.90ms, respectively). However, these figures are reversed when using shuffling.
Fully shuffled data sets are less sensitive to viewpoint changes with \textbf{GL\_POINTS} 
(30.71ms vs. 39.52ms $\rightarrow \times1.29$), but increase view dependence and incur overall negative impact on the performance of our more elaborate compute-based methods, including high-quality shading variants.
In opposition to these varied results, shuffled Morton order yields peak or close-to-peak performance over all rendering methods, regardless of viewpoint.
However, the best results for a particular combination of rendering method and point ordering were observed with the \textbf{dedup} approach (3.06ms vs. 3.04ms) on Morton sorted data sets, showing a margin of just 3\% between the close-up and overview scenario. %Another interesting observation is that the just-set method, which we expected to be the fastest in all cases, is slower than early-z in one case -- when the data set is shuffled and the user zooms out. We attribute this behavior to the overall higher payoff of early-z for shuffled data sets, where a few additional unsynchronized reads are economized by a lower number of distributed (unsynchronized) writes. 
A detailed illustration of the exemplary viewpoint benchmarks with different levels of zoom on an RTX 3090 is provided in Table~\ref{tab:scenarios_viewpoint} in the paper appendix.

\subsection{The Impact of Vertex Order on Performance}
\label{sec:ordering}
Tables~\ref{tab:scenarios} and~\ref{tab:scenarios_viewpoint} illustrate the impact of the four assessed vertex orders across different models and viewpoints. Elaborate compute-based methods perform better with Morton ordered layouts, \textbf{GL\_POINTS} and \textbf{busy-loop} perform better with shuffled points. Morton order predictably works well for methods that exploit locality, like \textbf{reduce} and \textbf{dedup} as well as close-up views with \textbf{GL\_POINTS} and \textbf{atomicMin}. Shuffled points are favorable for zoomed-out views with \textbf{GL\_POINTS} (3.9$\times$ over original) and \textbf{atomicMin} (2.1$\times$ over original). In contrast, shuffled Morton order is mostly view-independent and yields the highest performance for our compute-based methods in 65\% of all evaluated scenarios, outperformed only occasionally by Morton order with \textbf{dedup} or \textbf{reduce\&early-z} methods. It also raises the performance of high-quality shading by 7\% on average. 

The most drastic effects of choosing shuffled Morton, however, are observed for \textbf{GL\_POINTS}: rearranging input vertices with the new layout accelerates rendering performance by up to 4$\times$ (2.25$\times$ on average) on an RTX 3090 and up to 2.5$\times$ (1.6$\times$ on average) on an RTX 2070, compared to the original ordering. Creating custom vertex orders with the described policies only requires a one-time preprocessing step, which may take from less than a milliseconds to a few seconds, depending on the implementation and scene size. 
Hence, we claim that shuffled Morton is well-suited for the hardware pipeline and can be considered as a drop-in solution to accelerate applications based on \enum{GL\_POINTS}.

\subsection{Level-of-Detail Performance}
\label{sec:lod}

\begin{center}
\begin{table}
    \begin{tabular*}{\columnwidth}{@{\extracolsep{\fill}} rrrr}
    \toprule
    \#nodes   & points/node     & GL\_POINTS & compute \\ 
    \midrule
    1624     &   2.4k          &  718 FPS &  500 FPS    \\  
      80     &  49.4k          & 1615 FPS & 2096 FPS    \\  
    \end{tabular*}
    \caption{Performance of rendering 4 million points in LOD structures with different granularity settings. Compute-based methods benefit from fewer nodes with a more points.}
    \label{tab:perf_lod}
\end{table}
\end{center}

The \textit{layered point-cloud} (LPC)~\cite{LPC} structure and its variations are a widely used LOD structure for point clouds (e.g., \cite{QGIS,Potree,Entwine,Arena4D}) that stores subsamples of the full point cloud in the nodes of a spatial acceleration structure (e.g., binary tree, octree, kd-tree). Each node is essentially a small point cloud containing 100 to 20\,000 points, and the union of all smaller point clouds yields the original. 
%Even if the point clouds comprise hundreds of billions of points, 
This way, the number of points that are rendered in each frame can be limited to approximately 1 to 10 million points, which corresponds to about 100 to 4\,000 octree nodes. The points in each node are stored in separate vertex buffers, and in each frame the point cloud is rendered by invoking as many \code{glDrawArrays} calls as there are visible nodes.

% \begin{figure*}
%     \centering
%     \begin{subfigure}[t]{0.245\textwidth}
%         \includegraphics[width=\textwidth]{images/lod_0.jpg}
%         \caption{Level 0}
%     \end{subfigure}
%     \hfill
%     \begin{subfigure}[t]{0.245\textwidth}
%         \includegraphics[width=\textwidth]{images/lod_1.jpg}
%         \caption{Level 0-1}
%     \end{subfigure}
%     \hfill
%     \begin{subfigure}[t]{0.245\textwidth}
%         \includegraphics[width=\textwidth]{images/lod_2.jpg}
%         \caption{Level 0-2}
%     \end{subfigure}
%     \hfill
%     \begin{subfigure}[t]{0.245\textwidth}
%         \includegraphics[width=\textwidth]{images/lod_all.jpg}
%         \caption{All levels}
%     \end{subfigure}
%     \caption{An LOD structure that organizes all points into an octree. Each octree node is a small point cloud with a level-dependent density. Lion data set: 4 million points distributed in 1624 octree nodes, akin to 1624 point clouds with an average of about 2.4k points, each.}
%     \label{fig:lod}
% \end{figure*}

In order to evaluate the suitability of compute-based rendering for layered point clouds, we use PotreeConverter 2.1~\cite{PotreeConverter} to build an octree from the lion data set. With default granularity settings, PotreeConverter produces 1624 octree nodes with an average of 2.4k points per node. We also evaluate a modified case with lower granularity settings, which leads to an octree with 80 nodes and an average of 49.4k points in each node. Table~\ref{tab:perf_lod} shows the performances obtained by rendering both octree data sets with \textbf{GL\_POINTS} and \textbf{reduce\&early-z} as a representative for compute-based methods. Clearly, \textbf{GL\_POINTS} performs better when rendering a large number of nodes with a small amount of points in each, while compute performs better for fewer nodes with a higher number of points.
According to these results, we claim that compute-based rendering is useful to LOD rendering approaches that ensure individual LOD chunks with at least several ten thousand points. Methods that pack all points into a single buffer~\cite{Dachsbacher2003,schuetz-2019-CLOD} are very likely to benefit since compute approaches outperform \textbf{GL\_POINTS} in all evaluated orderings.

\subsection{Recommendations}

Interpreting the observations detailed in Table~\ref{tab:scenarios} and in the paper appendix (Tables~\ref{tab:scenarios_viewpoint},~\ref{tab:detailed_models_3090},~\ref{tab:detailed_viewpoints_3090},~\ref{tab:benchmark_2070} and~\ref{tab:benchmark_1060}), our recommendations for vertex order and rendering method are as follows:

\paragraph*{Recommended Vertex Order} We recommend using the \textbf{shuffled Morton} order for the brute-force rendering of large point clouds, since it is typically the fastest or within small margins of the fastest ordering according to our benchmarks, regardless of the technique being used. Regular Morton order can achieve peak performance with advanced compute shader approaches on modern GPU hardware, but is vastly outperformed by shuffled Morton order when rendering point clouds using the \enum{GL\_POINTS} primitive. 

\paragraph*{Recommended Rendering Method} The choice of the optimal rendering method is governed by the capabilities of the available hardware and graphics API. In our experiments, we observed peak performance results with the \textbf{dedup} approach in combination with Morton ordered points. However, these results strongly depend on recent GLSL extensions that may be unavailable for less specialized standards (e.g., WebGPU) and hardware-accelerated features of the NVIDIA Volta architecture or later. In contrast, the \textbf{reduce\&early-z} approach may be implemented with vendor-agnostic extensions, performs well on the older Pascal architecture (Table~\ref{tab:benchmark_1060} in appendix), and is consistently within slim performance margins of the \textbf{dedup} approach on newer models. We would therefore recommend using \textbf{reduce\&early-z} for wider support, and consider \textbf{dedup} as an optional improvement on Volta, Turing or Ampere architectures, since \textbf{dedup} shows consistently high performance with the smallest fluctuations between different viewpoints.

\section{Conclusions}

We have explored various methods for rendering point clouds with compute shaders, and shown that combining an \code{atomicMin}-based approach with early-z and warp-wide reduction or deduplication gives us the fastest results---up to an order of magnitude faster than using \enum{GL\_POINTS}. \textbf{dedup}, in particular, achieves consistently high performances in combination with Morton order, and a throughput of up to 50 billion points per second (800GB/s) on an RTX 3090, which amounts to 85\% of its specified memory bandwidth of 936GB/s. Additionally, a high-quality compute shader that computes the average of overlapping points was also shown to be faster than \enum{GL\_POINTS}, and results in a quality that is comparable to mipmapping or anisotropic filtering for textured meshes. \enum{GL\_POINTS} still outperforms compute when rendering a large number of small vertex buffers (up to a few thousands of points per buffer), but compute takes over when the buffers are larger (tens of thousands of points per buffer). From these results, we conclude that compute-based rendering with early-z and reduction enabled is a good choice for rendering unstructured point clouds with millions to hundreds of millions of points, and potentially useful for LOD rendering methods that employ larger LOD chunks. 

We also investigated the impact of vertex order on rendering performance and found that Morton order and shuffled vertex buffers outperform each other in different scenarios. Combining both into a shuffled Morton vertex order preserves the advantages of both, and yields consistently high rendering performance over various different viewpoints. 

In the future, we would like to further explore the application of compute shaders to the LOD rendering of point clouds. We believe that with sophisticated work aggregation schemes, compute shaders could prove advantageous for the fast and fine-grained selection and rendering of the most suitable LOD chunks, compared to the current layered point cloud applications that select a set of coarse chunks on the CPU side and then invoke one draw call for each. 

The full source code and windows binaries are available at \url{https://github.com/m-schuetz/compute_rasterizer}.

\section{Acknowledgements}

The authors wish to thank \emph{Riegl Laser Measurement Systems} for providing the data set of the town of Retz, \emph{NVIDIA} for the Endeavor building site data set, \emph{Weiss AG} for the lifeboat data set, and \emph{TU Wien, Institute of History of Art, Building Archaeology and Restoration} for the Candi Sari data set.

This research has been funded by the FFG project \textit{LargeClouds2BIM} and FWF project no. P32418, as well as the Research Cluster “Smart Communities and Technologies (Smart CT)” at TU Wien.

\printbibliography

\clearpage

\appendix

\onecolumn

\section{Illustrated Benchmarks}

\begin{table}[H]
\begin{tabular*}{\textwidth}{ll}
\toprule
 \textbf{Model:} Retz (all points outside frustum) & \textbf{Points:} 145 million (2.3GB) \\ 
 %\midrule
 \includegraphics[width=0.45\textwidth]{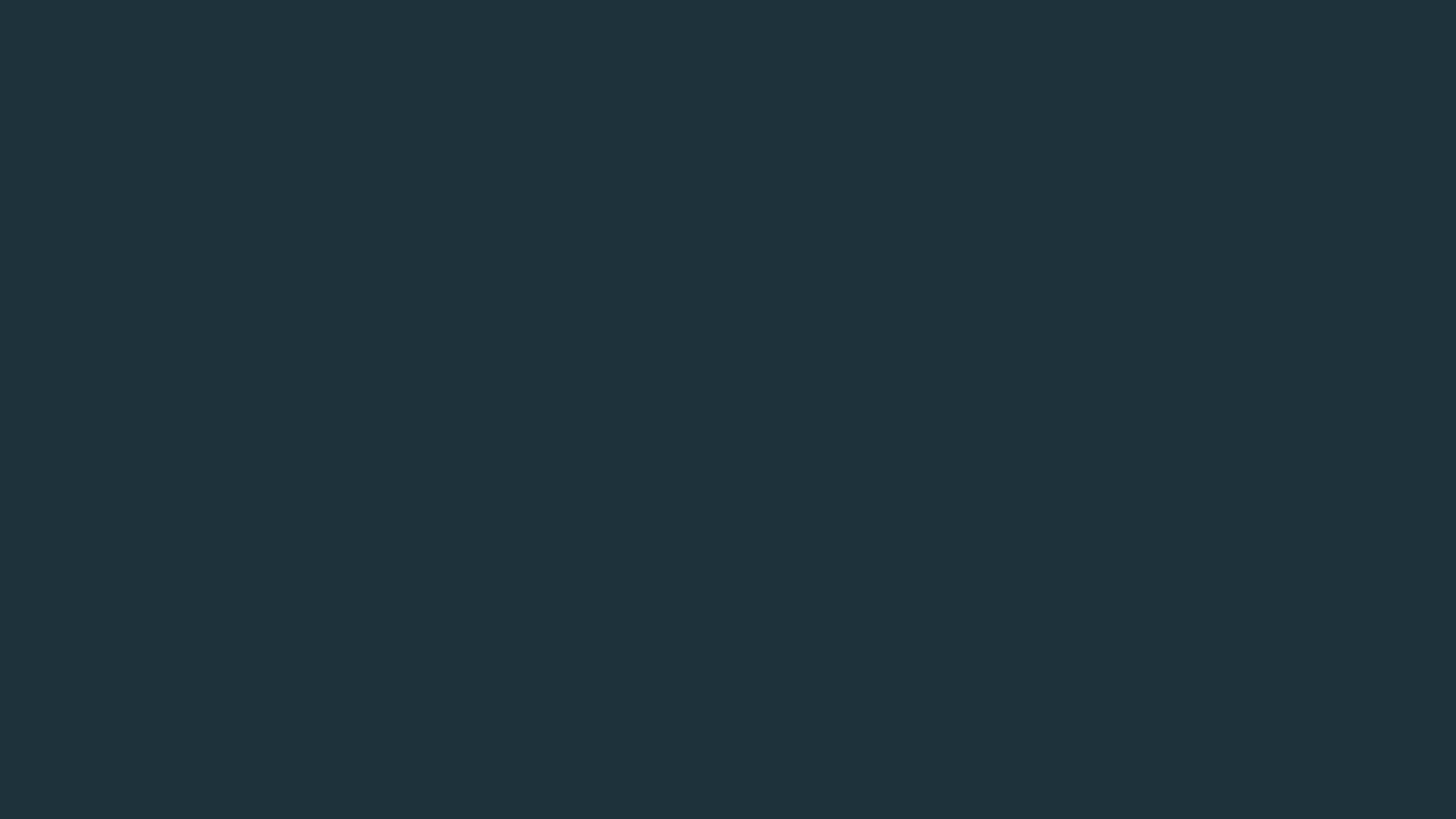} & 
 \includegraphics[width=0.5\textwidth]{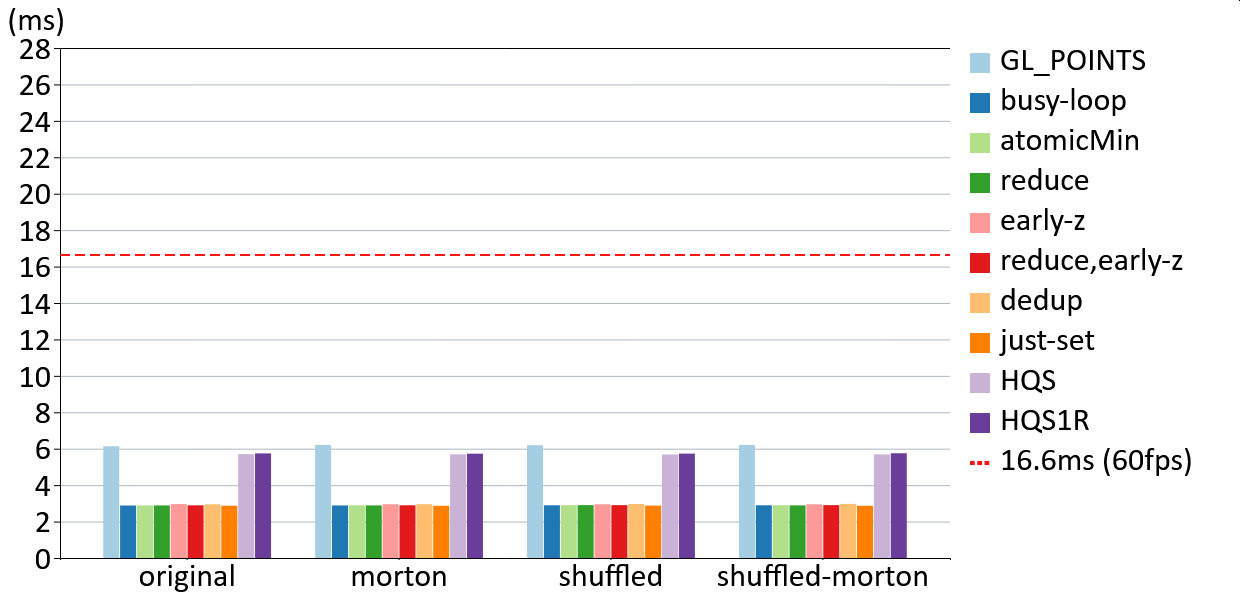} \\
 \midrule
 \textbf{Model:} Retz (closeup) & \textbf{Points:} 145 million (2.3GB) \\ 
 \includegraphics[width=0.45\textwidth]{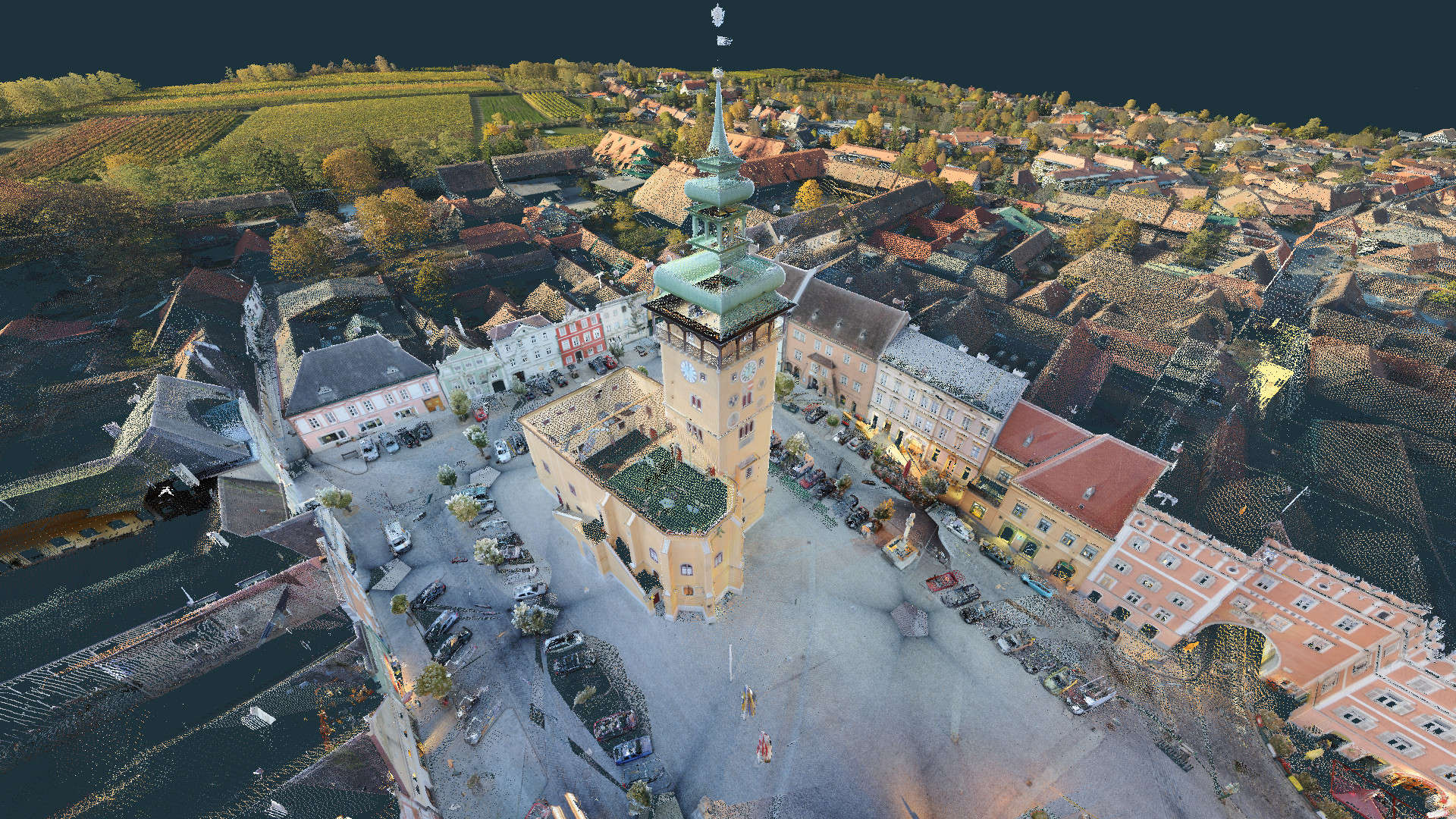} & 
 \includegraphics[width=0.5\textwidth]{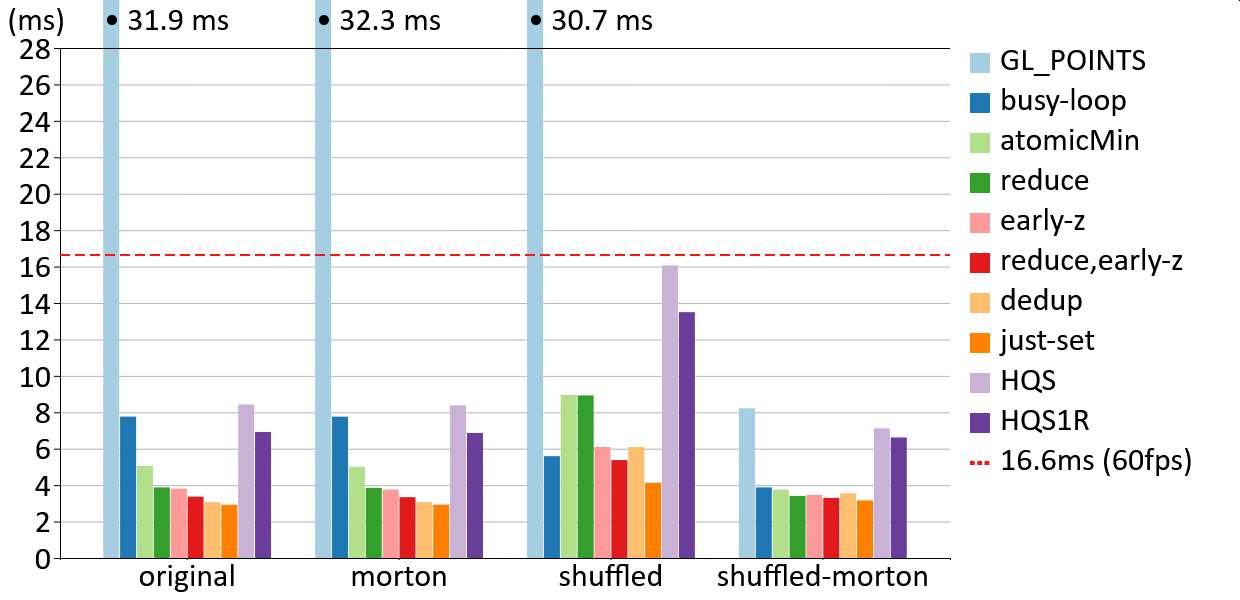} \\
 \midrule
 \textbf{Model:} Retz (overview) & \textbf{Points:} 145 million (2.3GB)  \\ 
 \includegraphics[width=0.45\textwidth]{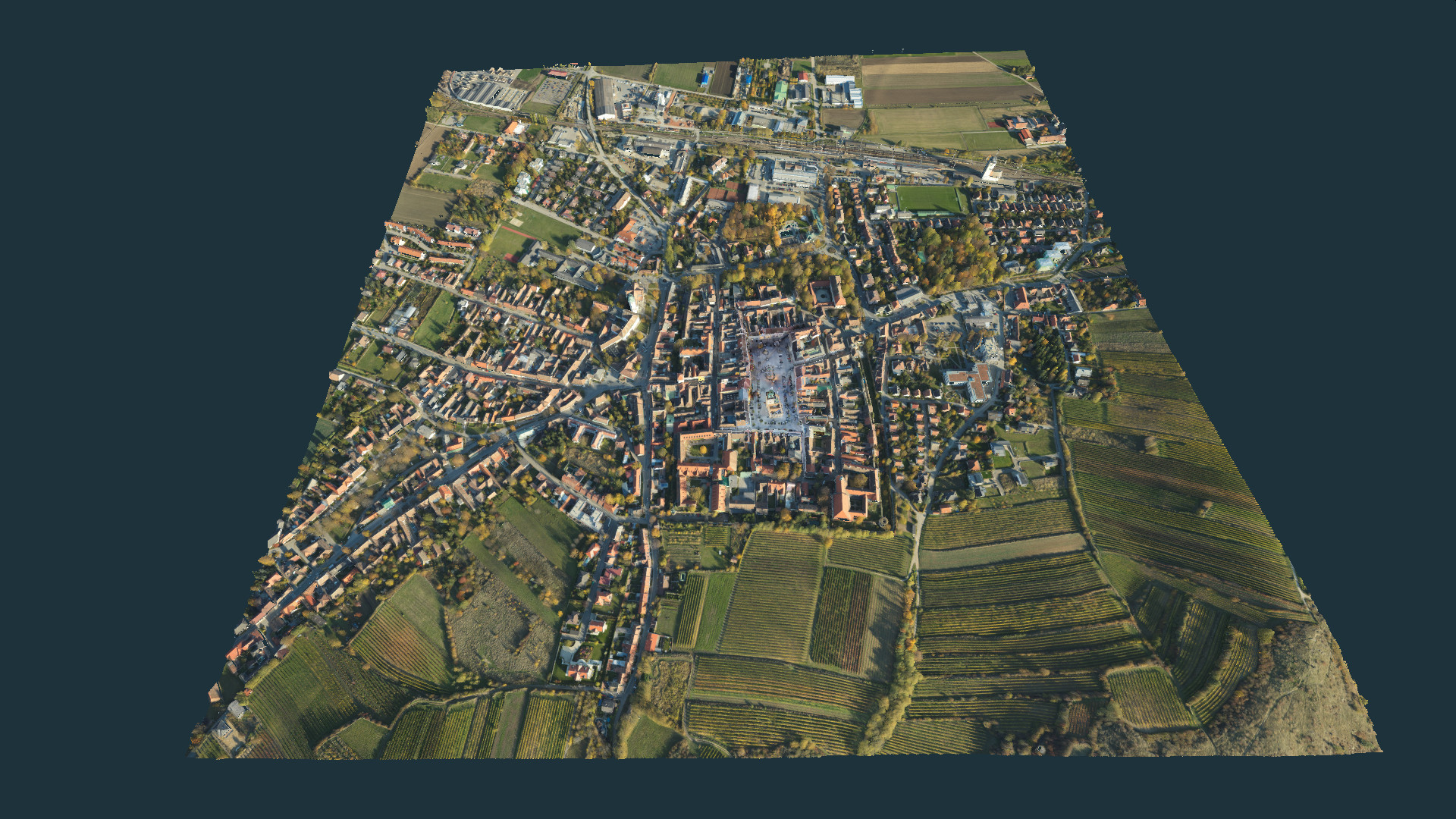} & 
 \includegraphics[width=0.5\textwidth]{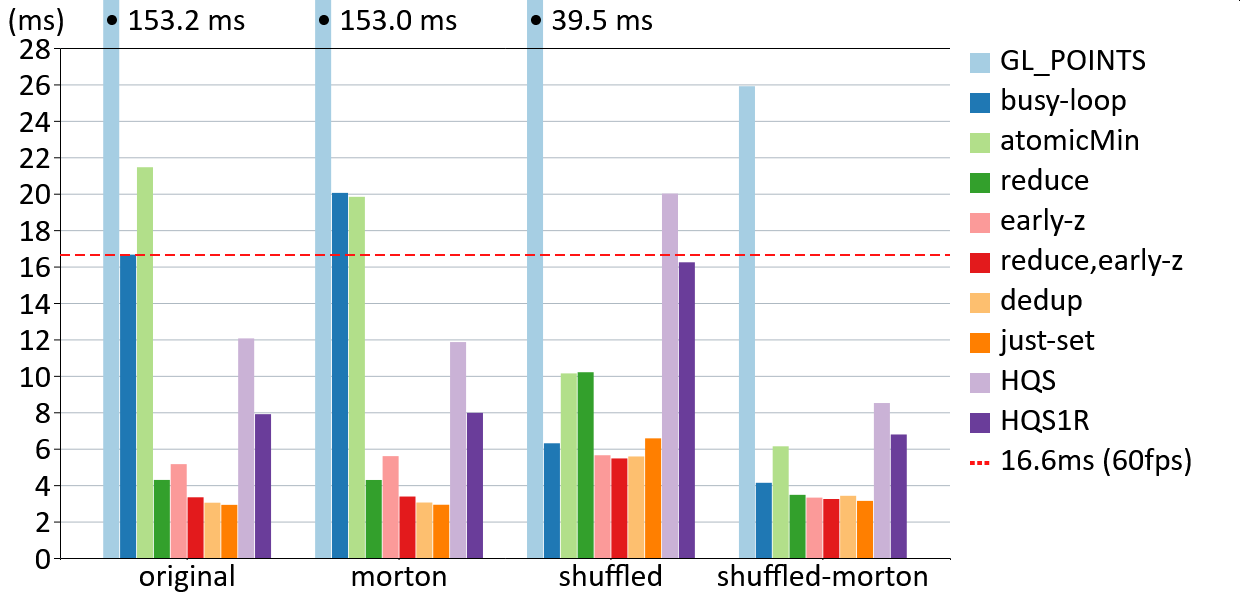} \\
\bottomrule
\end{tabular*}
\caption{Frame times (ms) for Retz in three different viewpoints on an RTX 3090 (lower is better). \enum{GL\_POINTS} and atomicMin are strongly affected by the viewpoint, while the combinations of reduce and either early-z or dedup are highly robust to changes of the viewpoint.}
\label{tab:scenarios_viewpoint}
\end{table}

\newpage

\section{Detailed Benchmark Results}

\definecolor{cSlowest}{RGB}{220,160,160}
\definecolor{cFastest}{RGB}{160,220,160}
\definecolor{cAlsoFast}{RGB}{220,255,220}

The following tables report the frame times for various models rendered on different GPUs using several methods. 
The fastest frame times in a row and methods within 5\% of the fastest are highlighted in dark green, and methods within 10\% of the fastest are highlighted in light green. All times in milliseconds. The set method is exempt from the calculations as it is not a practicable solution. 

The results are discussed in Section~\ref{sec:evaluation}.

\begin{table}[H]
\setlength\tabcolsep{5pt} % default value: 6pt
\begin{tabular}{|c|l|r|rrrrr|rr|rr|}
\hline
                          &                 &            & \multicolumn{5}{c|}{\textbf{Basic Compute}}                                                                                             & \multicolumn{2}{c|}{\textbf{Misc}}                    & \multicolumn{2}{c|}{\textbf{High-Quality}} \\
\thead{Model \\ (\#Points)}    &  Vertex Order   & GL & atomicMin & reduce & early-z & \thead{reduce, \\ early-z} & dedup & \thead{busy \\ loop} & set & HQS & \thead{HQS1R}      \\
\hline
\multirow{4}{*}{\thead{Lion \\ (4M)}}     &        original & \cellcolor{cSlowest}  0.59 & \cellcolor{cFastest}  0.28 & \cellcolor{cFastest}  0.28 & \cellcolor{cAlsoFast}  0.30 & \cellcolor{cAlsoFast}  0.30 & \cellcolor{cAlsoFast}  0.30 & \cellcolor{cSlowest}  0.58 &   0.25 &   0.53 &   0.52 \\
                          &          Morton & \cellcolor{cSlowest}  0.70 &   0.27 & \cellcolor{cFastest}  0.25 &   0.27 & \cellcolor{cAlsoFast}  0.27 & \cellcolor{cFastest}  0.24 &   0.46 
&   0.21 &   0.48 &   0.47 \\
                          &        shuffled & \cellcolor{cSlowest}  1.34 &   0.35 &   0.35 & \cellcolor{cFastest}  0.32 & \cellcolor{cFastest}  0.32 & \cellcolor{cFastest}  0.32 &   0.35 &   0.27 &   0.89 &   0.65 \\
                          & shuffled Morton & \cellcolor{cSlowest}  0.47 & \cellcolor{cAlsoFast}  0.28 & \cellcolor{cFastest}  0.26 & \cellcolor{cFastest}  0.27 & \cellcolor{cFastest}  0.27 & \cellcolor{cFastest}  0.26 &   0.38 &   0.21 & \cellcolor{cSlowest}  0.48 & \cellcolor{cSlowest}  0.48 \\
\hline
\multirow{4}{*}{\thead{Lifeboat \\ (47M)}} &        original & \cellcolor{cSlowest}  9.19 &   1.93 &   1.71 & \cellcolor{cAlsoFast}  1.29 & \cellcolor{cFastest}  1.27 & \cellcolor{cFastest}  1.23 &   2.62 
&   1.07 &   2.82 &   2.45 \\
                          &          Morton & \cellcolor{cSlowest} 12.29 &   2.14 &   1.60 &   1.50 &   1.30 & \cellcolor{cFastest}  1.13 &   3.35 &   1.06 &   2.91 &   2.56 \\
                          &        shuffled & \cellcolor{cSlowest} 15.00 &   5.16 &   5.14 & \cellcolor{cAlsoFast}  2.64 & \cellcolor{cFastest}  2.52 & \cellcolor{cAlsoFast}  2.61 & \cellcolor{cFastest}  2.42 &   1.82 &   7.25 &   5.67 \\
                          & shuffled Morton & \cellcolor{cSlowest}  5.03 &   1.65 &   1.38 & \cellcolor{cFastest}  1.27 & \cellcolor{cFastest}  1.26 & \cellcolor{cFastest}  1.24 &   2.06 &   1.14 &   2.56 &   2.48 \\
\hline
\multirow{4}{*}{\thead{Retz \\ (145M)}}     &        original & \cellcolor{cSlowest} 31.98 &   5.07 &   3.91 &   3.81 & \cellcolor{cAlsoFast}  3.36 & \cellcolor{cFastest}  3.07 &   7.66 &   2.93 &   8.40 & 
  6.87 \\
                          &          Morton & \cellcolor{cSlowest} 32.05 &   5.01 &   3.86 &   3.76 & \cellcolor{cAlsoFast}  3.35 & \cellcolor{cFastest}  3.06 &   7.65 &   2.93 &   8.36 & 
  6.83 \\
                          &        shuffled & \cellcolor{cSlowest} 30.57 &   8.58 &   8.70 &   6.10 & \cellcolor{cFastest}  5.35 &   5.93 & \cellcolor{cFastest}  5.55 &   4.24 &  16.23 &  
13.18 \\
                          & shuffled Morton & \cellcolor{cSlowest}  8.22 &   3.75 & \cellcolor{cFastest}  3.39 & \cellcolor{cFastest}  3.46 & \cellcolor{cFastest}  3.31 & \cellcolor{cFastest}  3.44 &   3.84 &   3.18 &   7.08 &   6.61 \\
\hline
\multirow{4}{*}{\thead{Endeavor \\ (796M)}} &        original & \cellcolor{cSlowest} 71.62 &  26.13 &  25.79 & \cellcolor{cAlsoFast} 19.14 & \cellcolor{cFastest} 17.86 & \cellcolor{cAlsoFast} 19.00 &  21.16 &  20.17 &  40.35 &  37.27 \\
                          &          Morton & \cellcolor{cSlowest}131.49 &  36.09 &  21.53 & \cellcolor{cAlsoFast} 16.56 & \cellcolor{cFastest} 15.86 & \cellcolor{cFastest} 15.26 &  22.84 
&  15.05 &  35.34 &  31.57 \\
                          &        shuffled & \cellcolor{cSlowest}162.37 &  43.11 &  44.13 &  28.72 & \cellcolor{cAlsoFast} 25.62 &  27.93 & \cellcolor{cFastest} 24.30 &  22.09 &  54.06 & 
 50.51 \\
                          & shuffled Morton & \cellcolor{cSlowest} 35.62 &  17.84 & \cellcolor{cFastest} 16.62 & \cellcolor{cAlsoFast} 17.68 & \cellcolor{cFastest} 16.20 &  18.01 & \cellcolor{cFastest} 16.56 &  16.23 &  31.50 &  31.46 \\
\hline
\end{tabular}
\caption{Frame times (ms) of four models using several different rendering methods. GPU: \textbf{RTX 3090}. This table is basis for the charts in Table~\ref{tab:scenarios}.}
\label{tab:detailed_models_3090}
\end{table}

\begin{table}[H]
\setlength\tabcolsep{5pt} % default value: 6pt
\begin{tabular}{|c|l|r|rrrrr|rr|rr|}
\hline
                          &                 &            & \multicolumn{5}{c|}{\textbf{Basic Compute}}                                                                                             & \multicolumn{2}{c|}{\textbf{Misc}}                    & \multicolumn{2}{c|}{\textbf{High-Quality}} \\
\thead{Model \\ (Viewpoint)}    &  Vertex Order   & GL & atomicMin & reduce & early-z & \thead{reduce, \\ early-z} & dedup & \thead{busy \\ loop} & set & HQS & \thead{HQS1R}      \\
\hline
\multirow{4}{*}{\thead{Retz \\ (no point \\ in frustum)}} &        original & \cellcolor{cSlowest}  6.13 & \cellcolor{cFastest}  2.89 & \cellcolor{cFastest}  2.89 & \cellcolor{cFastest}  2.96 & \cellcolor{cFastest}  2.89 & \cellcolor{cFastest}  2.95 & \cellcolor{cFastest}  2.88 &   2.87 &   5.70 &   5.74 \\
                          &          Morton & \cellcolor{cSlowest}  6.20 & \cellcolor{cFastest}  2.90 & \cellcolor{cFastest}  2.89 & \cellcolor{cFastest}  2.95 & \cellcolor{cFastest}  2.89 & \cellcolor{cFastest}  2.96 & \cellcolor{cFastest}  2.89 &   2.87 &   5.69 &   5.73 \\
                          &        shuffled & \cellcolor{cSlowest}  6.19 & \cellcolor{cFastest}  2.90 & \cellcolor{cFastest}  2.90 & \cellcolor{cFastest}  2.95 & \cellcolor{cFastest}  2.90 & \cellcolor{cFastest}  2.96 & \cellcolor{cFastest}  2.89 &   2.88 &   5.68 &   5.73 \\
                          & shuffled Morton & \cellcolor{cSlowest}  6.21 & \cellcolor{cFastest}  2.90 & \cellcolor{cFastest}  2.89 & \cellcolor{cFastest}  2.95 & \cellcolor{cFastest}  2.90 & \cellcolor{cFastest}  2.97 & \cellcolor{cFastest}  2.90 &   2.87 &   5.69 &   5.75 \\
\hline
\multirow{4}{*}{\thead{Retz \\ (closeup)}} &        original & \cellcolor{cSlowest} 31.95 &   5.07 &   3.88 &   3.81 & \cellcolor{cAlsoFast}  3.37 & \cellcolor{cFastest}  3.06 &   7.76 &   2.92 &   8.43 &   6.91 \\
                          &          Morton & \cellcolor{cSlowest} 32.26 &   5.01 &   3.85 &   3.76 & \cellcolor{cAlsoFast}  3.34 & \cellcolor{cFastest}  3.07 &   7.76 &   2.93 &   8.38 & 
  6.86 \\
                          &        shuffled & \cellcolor{cSlowest} 30.71 &   8.95 &   8.93 &   6.10 & \cellcolor{cFastest}  5.37 &   6.09 & \cellcolor{cFastest}  5.59 &   4.13 &  16.06 &  
13.50 \\
                          & shuffled Morton & \cellcolor{cSlowest}  8.22 &   3.75 & \cellcolor{cFastest}  3.41 & \cellcolor{cFastest}  3.46 & \cellcolor{cFastest}  3.30 & \cellcolor{cAlsoFast}  3.55 &   3.88 &   3.16 &   7.12 &   6.62 \\
\hline
\multirow{4}{*}{\thead{Retz \\ (overview)}} &        original & \cellcolor{cSlowest}153.22 &  21.45 &   4.29 &   5.15 & \cellcolor{cAlsoFast}  3.33 & \cellcolor{cFastest}  3.04 &  16.66 &   2.92 &  12.06 &   7.90 \\
                          &          Morton & \cellcolor{cSlowest}152.96 &  19.83 &   4.28 &   5.59 &   3.37 & \cellcolor{cFastest}  3.05 &  20.05 &   2.92 &  11.85 &   7.97 \\
                          &        shuffled & \cellcolor{cSlowest} 39.52 &  10.14 &  10.20 & \cellcolor{cFastest}  5.64 & \cellcolor{cFastest}  5.47 & \cellcolor{cFastest}  5.57 &   6.30 &   6.57 &  20.01 &  16.23 \\
                          & shuffled Morton & \cellcolor{cSlowest} 25.90 &   6.13 & \cellcolor{cAlsoFast}  3.47 & \cellcolor{cFastest}  3.32 & \cellcolor{cFastest}  3.24 & \cellcolor{cAlsoFast}  3.42 &   4.13 &   3.14 &   8.51 &   6.78 \\
\hline
\end{tabular}
\caption{Frame times (ms) of the Retz model (145M points) in three different viewpoints on an \textbf{RTX 3090}. This table is basis for the charts in Table~\ref{tab:scenarios_viewpoint}.}
\label{tab:detailed_viewpoints_3090}
\end{table}

\begin{table}[H]
\setlength\tabcolsep{5pt} % default value: 6pt
\begin{tabular}{|c|l|r|rrrrr|rr|rr|}
\hline
                          &                 &            & \multicolumn{5}{c|}{\textbf{Basic Compute}}                                                                                             & \multicolumn{2}{c|}{\textbf{Misc}}                    & \multicolumn{2}{c|}{\textbf{High-Quality}} \\
\thead{Model \\ (\#Points)}    &  Vertex Order   & GL & atomicMin & reduce & early-z & \thead{reduce, \\ early-z} & dedup & \thead{busy \\ loop} & set & HQS & \thead{HQS1R}      \\
\hline
\multirow{4}{*}{\thead{Lion \\ (4M)}}     &        original &   0.78 & \cellcolor{cFastest}  0.51 & \cellcolor{cFastest}  0.50 & \cellcolor{cAlsoFast}  0.53 & \cellcolor{cFastest}  0.52 & \cellcolor{cAlsoFast}  0.54 & \cellcolor{cSlowest}  0.92 &   0.37 & \cellcolor{cSlowest}  0.90 & \cellcolor{cSlowest}  0.93 \\
                          &          Morton & \cellcolor{cSlowest}  1.10 &   0.52 & \cellcolor{cAlsoFast}  0.46 &   0.48 & \cellcolor{cAlsoFast}  0.46 & \cellcolor{cFastest}  0.44 &   0.82 &   0.35 &   0.87 &   0.89 \\
                          &        shuffled &   1.85 &   1.12 &   1.12 &   0.82 & \cellcolor{cFastest}  0.72 & \cellcolor{cFastest}  0.72 &   0.87 &   0.45 & \cellcolor{cSlowest}  2.61 &  
 2.11 \\
                          & shuffled Morton &   0.85 &   0.55 & \cellcolor{cAlsoFast}  0.51 & \cellcolor{cFastest}  0.50 & \cellcolor{cFastest}  0.49 & \cellcolor{cFastest}  0.49 &   0.79 
&   0.38 &   0.93 & \cellcolor{cSlowest}  0.99 \\
\hline
\multirow{4}{*}{\thead{Lifeboat \\ (47M)}} &        original & \cellcolor{cSlowest} 11.26 &   3.66 &   3.11 & \cellcolor{cFastest}  2.49 & \cellcolor{cFastest}  2.47 & \cellcolor{cFastest}  2.44 &   4.33 &   2.11 &   5.21 &   4.72 \\
                          &          Morton & \cellcolor{cSlowest} 16.57 &   4.43 &   3.21 &   3.12 &   2.70 & \cellcolor{cFastest}  2.36 &   6.01 &   2.10 &   5.52 &   4.98 \\
                          &        shuffled &  21.68 & \cellcolor{cSlowest} 22.26 & \cellcolor{cSlowest} 22.96 &   9.67 &   9.22 &   9.93 & \cellcolor{cFastest}  7.82 &   8.50 & \cellcolor{cSlowest} 22.83 &  20.70 \\
                          & shuffled Morton & \cellcolor{cSlowest}  8.96 &   4.28 &   3.39 & \cellcolor{cFastest}  3.00 & \cellcolor{cFastest}  3.00 & \cellcolor{cFastest}  2.98 &   5.02 &   2.62 &   5.91 &   5.89 \\
\hline
\multirow{4}{*}{\thead{Retz \\ (145M)}}     &        original & \cellcolor{cSlowest} 42.10 &   9.73 &   7.50 &   7.93 &   7.00 & \cellcolor{cFastest}  6.33 &  13.83 &   5.89 &  15.75 &  13.50 \\
                          &          Morton & \cellcolor{cSlowest} 42.80 &   9.71 &   7.48 &   7.91 & \cellcolor{cAlsoFast}  6.92 & \cellcolor{cFastest}  6.31 &  13.79 &   5.90 &  15.89 & 
 13.50 \\
                          &        shuffled &  45.63 &  37.55 &  38.73 &  18.30 &  18.73 &  17.90 & \cellcolor{cFastest} 16.15 &  16.53 & \cellcolor{cSlowest} 51.23 &  44.32 \\
                          & shuffled Morton & \cellcolor{cSlowest} 16.16 &   9.20 & \cellcolor{cAlsoFast}  8.14 & \cellcolor{cFastest}  7.70 & \cellcolor{cFastest}  7.65 & \cellcolor{cFastest}  7.60 &   9.25 &   7.29 & \cellcolor{cSlowest} 15.81 & \cellcolor{cSlowest} 15.43 \\
\hline
\end{tabular}
\caption{Frame times (ms) using several different rendering methods on an \textbf{RTX 2070}.}
\label{tab:benchmark_2070}
\end{table}

\begin{table}[H]
\setlength\tabcolsep{5pt} % default value: 6pt
\begin{tabular}{|c|l|r|rrrrr|rr|rr|}
\hline
                          &                 &            & \multicolumn{5}{c|}{\textbf{Basic Compute}}                                                                                             & \multicolumn{2}{c|}{\textbf{Misc}}                    & \multicolumn{2}{c|}{\textbf{High-Quality}} \\
\thead{Model \\ (\#Points)}    &  Vertex Order   & GL & atomicMin & reduce & early-z & \thead{reduce, \\ early-z} & dedup & \thead{busy \\ loop} & set & HQS & \thead{HQS1R}      \\
\hline
\multirow{4}{*}{\thead{Lion \\ (4M)}}     &        original &   2.08 & \cellcolor{cFastest}  1.13 & \cellcolor{cFastest}  1.15 & \cellcolor{cFastest}  1.18 & \cellcolor{cFastest}  1.16 &   1.72 &   1.87 &   1.00 &   2.11 & \cellcolor{cSlowest}  2.23 \\
                          &          Morton & \cellcolor{cSlowest}  2.53 & \cellcolor{cAlsoFast}  1.13 & \cellcolor{cFastest}  1.05 & \cellcolor{cFastest}  1.09 & \cellcolor{cFastest}  1.06 & \cellcolor{cAlsoFast}  1.13 &   1.81 &   0.81 &   2.09 &   2.13 \\
                          &        shuffled &   4.89 &   4.60 &   4.61 & \cellcolor{cFastest}  3.09 & \cellcolor{cFastest}  3.00 &   3.32 &   3.59 &   1.89 & \cellcolor{cSlowest}  8.33 &  
 7.51 \\
                          & shuffled Morton &   2.16 &   1.29 & \cellcolor{cAlsoFast}  1.21 & \cellcolor{cFastest}  1.16 & \cellcolor{cFastest}  1.15 & \cellcolor{cAlsoFast}  1.24 &   1.71 &   0.87 & \cellcolor{cSlowest}  2.33 & \cellcolor{cSlowest}  2.37 \\
\hline
\multirow{4}{*}{\thead{Lifeboat \\ (47M)}} &        original & \cellcolor{cSlowest} 24.17 &   7.76 &   7.14 & \cellcolor{cFastest}  6.13 & \cellcolor{cFastest}  6.09 &   9.84 &   9.29 &   5.61 &  12.48 &  
11.95 \\
                          &          Morton & \cellcolor{cSlowest} 34.05 &   8.60 & \cellcolor{cAlsoFast}  6.92 & \cellcolor{cAlsoFast}  6.97 & \cellcolor{cFastest}  6.42 & \cellcolor{cFastest}  6.62 &  13.05 &   5.11 &  13.50 &  12.47 \\
                          &        shuffled & \cellcolor{cSlowest} 84.06 &  62.45 &  63.04 & \cellcolor{cAlsoFast} 32.07 & \cellcolor{cFastest} 31.01 &  35.68 & \cellcolor{cFastest} 30.14 
&  26.00 &  67.58 &  68.03 \\
                          & shuffled Morton & \cellcolor{cSlowest} 23.66 &   9.21 &   8.04 & \cellcolor{cFastest}  7.11 & \cellcolor{cFastest}  7.23 & \cellcolor{cAlsoFast}  7.70 &  11.03 
&   5.85 &  14.03 &  14.51 \\
\hline
\end{tabular}
\caption{Frame times (ms) using several different rendering methods on a \textbf{GTX 1060 (3GB)}.}
\label{tab:benchmark_1060}
\end{table}

\newpage
\section{High-Quality Rendering Visual Comparison}

Below, we provide visual comparisons of rendering tested scenes with and without the high-quality antialiasing shaders.

\begin{figure}[h!]
    \centering
    \begin{subfigure}{0.49\linewidth}
    \includegraphics[trim=430 170 420 160, clip,width=\linewidth]{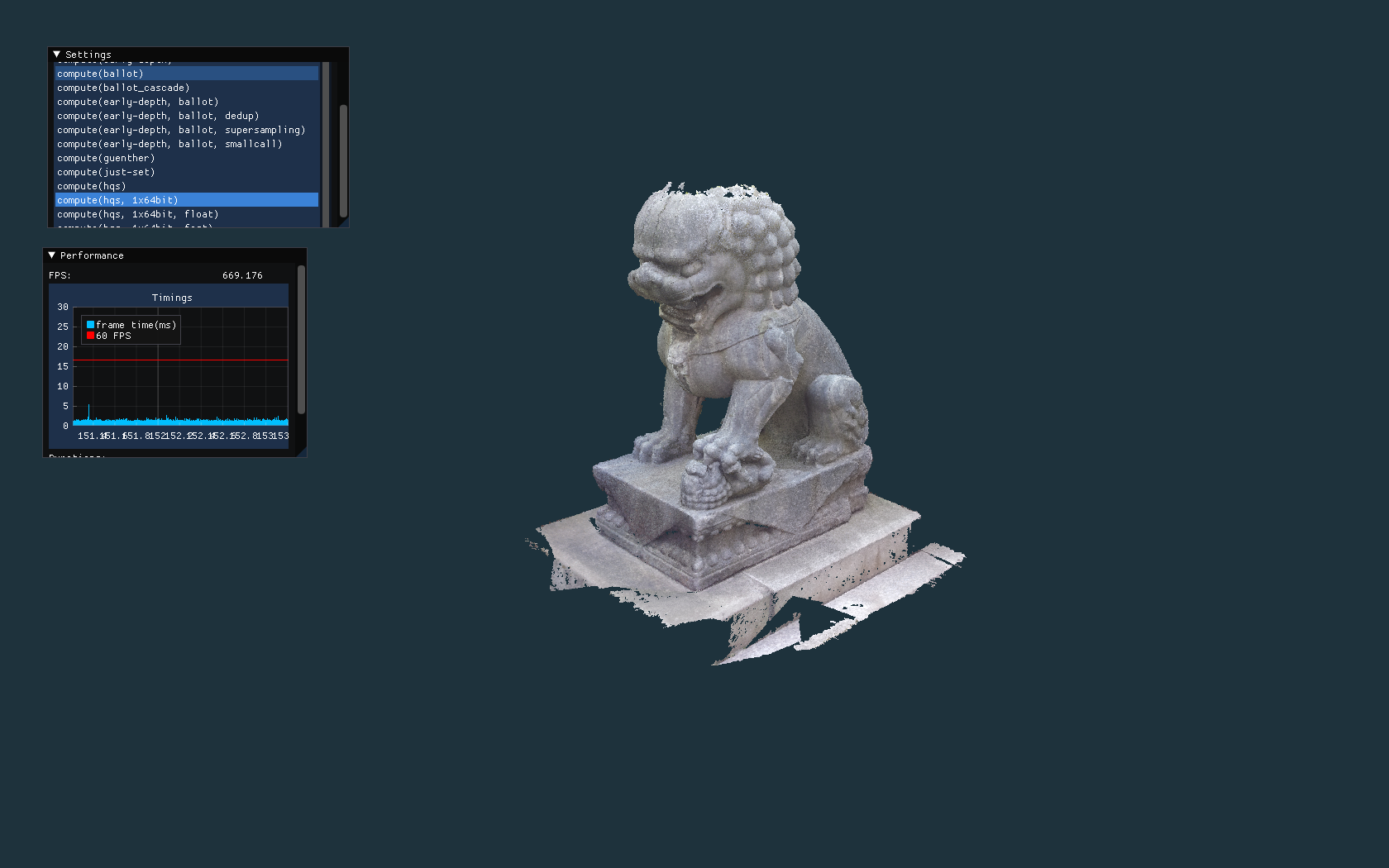}
    \begin{picture}(0,0)
    \put(2,215){\includegraphics[trim=570 555 630 172, clip,height=2.4cm]{images/im4}}
    \end{picture}
    \caption{Lion statue with close-up inset, rendered with \enum{GL\_POINTS}}
    \end{subfigure}
    \hfill
    \begin{subfigure}{0.49\linewidth}
    \includegraphics[trim=430 170 420 160, clip,width=\linewidth]{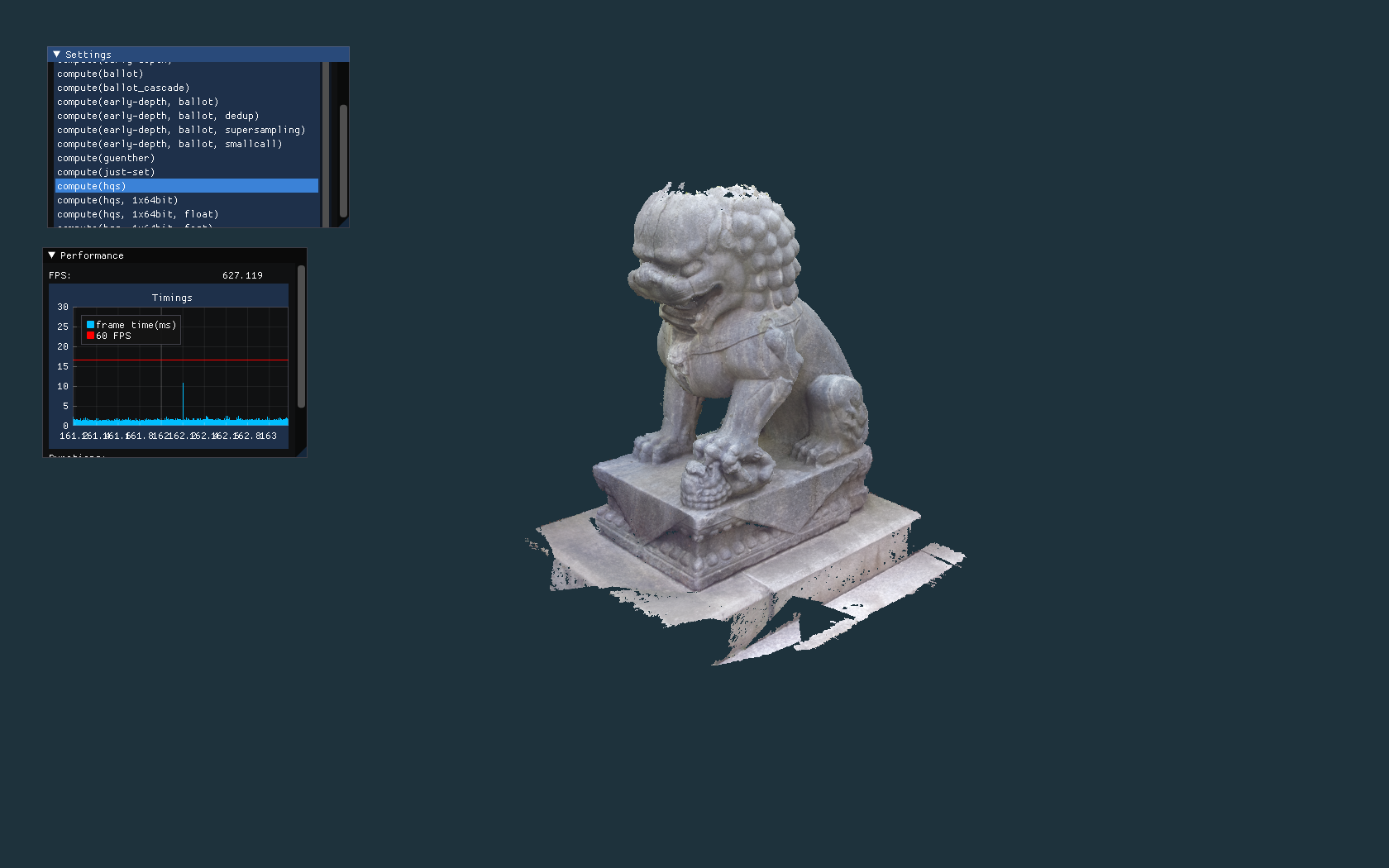}
    \begin{picture}(0,0)
    \put(2,215){\includegraphics[trim=570 555 630 172, clip,height=2.4cm]{images/im3}}
    \end{picture}
    \caption{Lion statue with close-up inset, rendered with \textbf{HQS1R}}
    \end{subfigure}
\end{figure}

\begin{figure}[h!]
    \centering
    \begin{subfigure}{0.49\linewidth}
   \includegraphics[trim=370 80 370 185, clip,width=\linewidth]{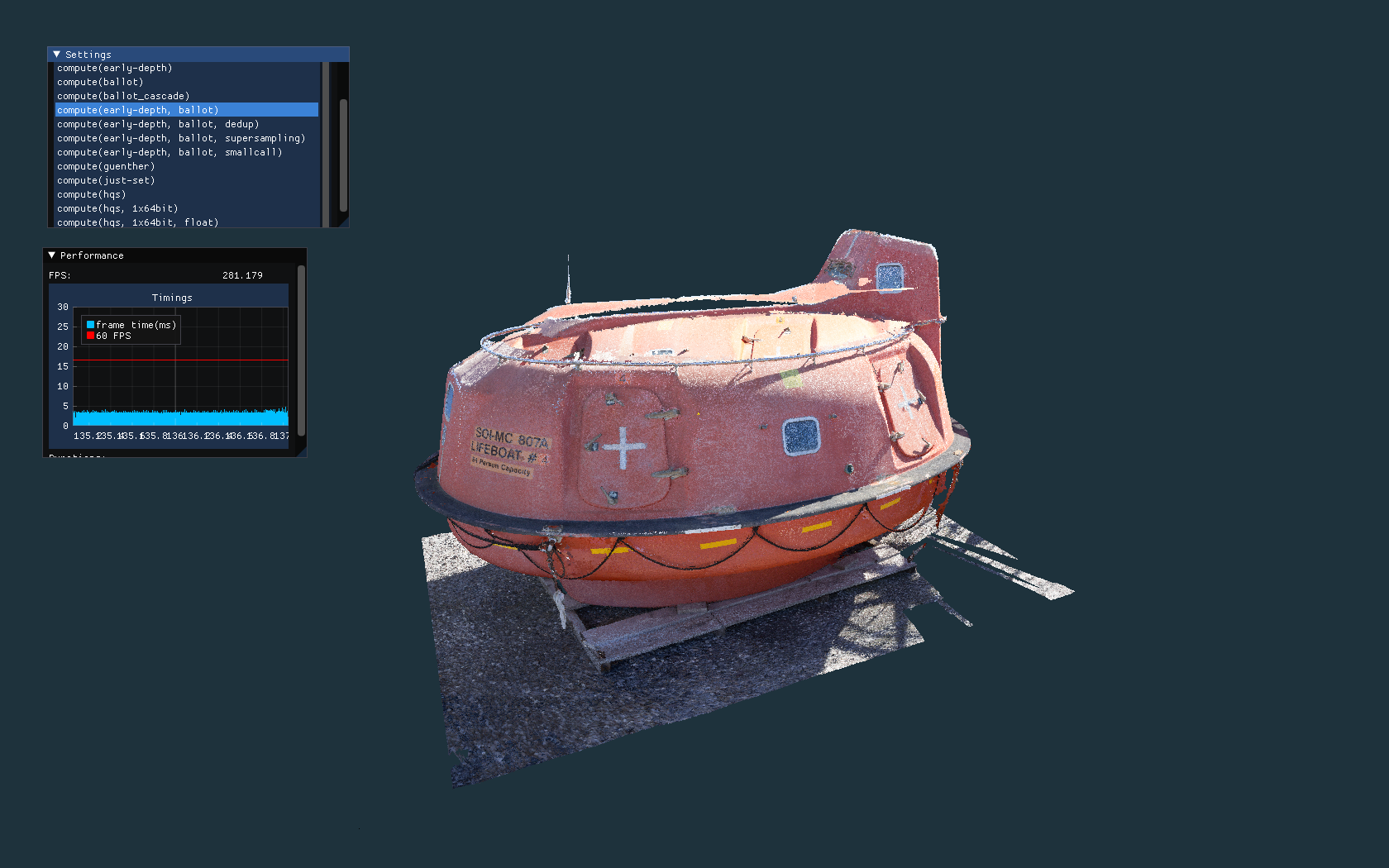}
    \begin{picture}(0,0)
    \put(177,13){\includegraphics[trim=703 370 510 370, clip,height=2.4cm]{images/img1}}
    \end{picture}
    \caption{Lifeboat  with close-up inset, rendered with \enum{GL\_POINTS}}
    \end{subfigure}
    \hfill
    \begin{subfigure}{0.49\linewidth}
    \includegraphics[trim=370 80 370 185, clip,width=\linewidth]{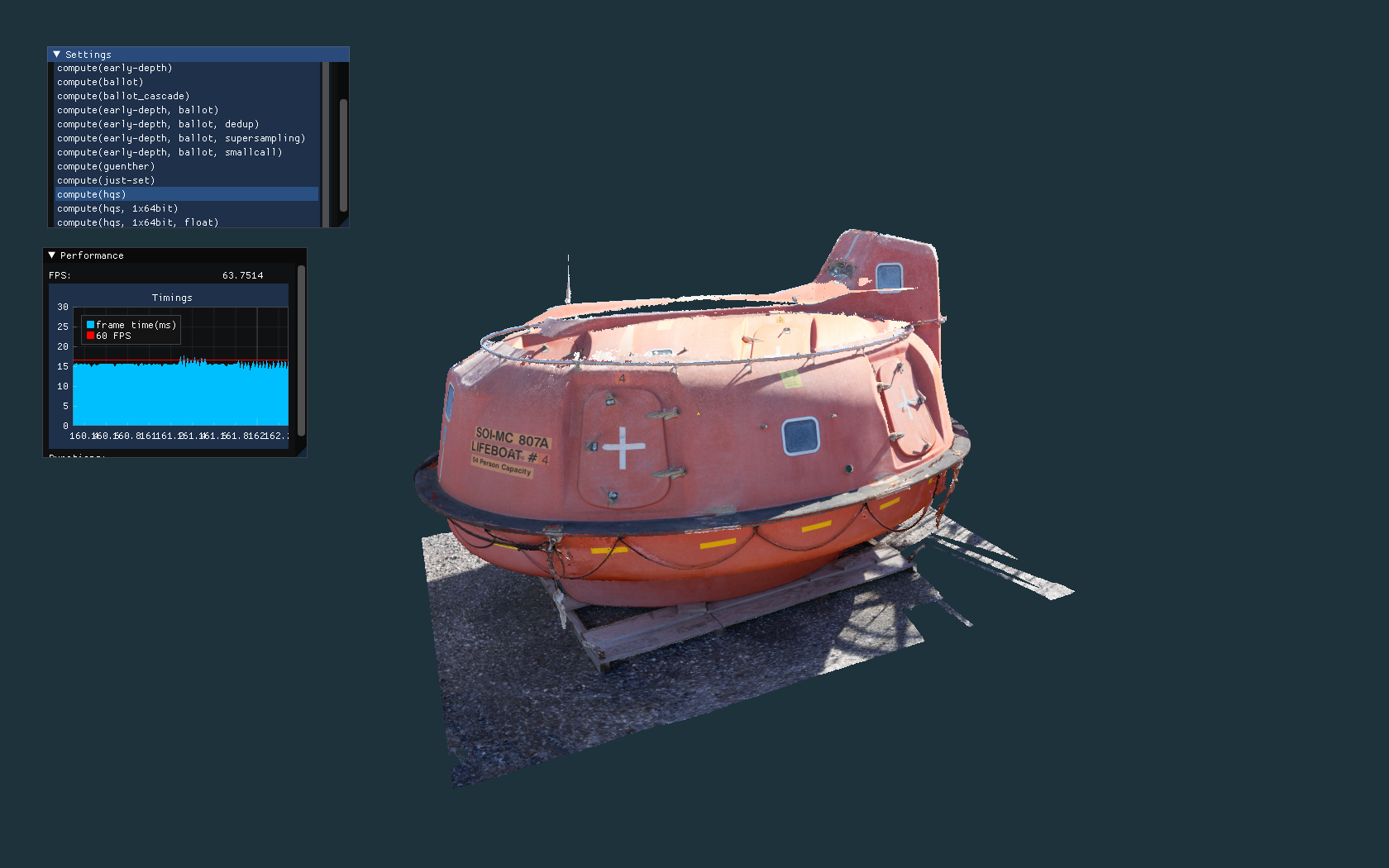}
    \begin{picture}(0,0)
    \put(177,13){\includegraphics[trim=703 370 510 370, clip,height=2.4cm]{images/img2}}
    \end{picture}
    \caption{Lifeboat with close-up inset, rendered with \textbf{HQS1R}}
    \end{subfigure}
\end{figure}

\begin{figure}[h!]
    \centering
    \begin{subfigure}{0.49\linewidth}
    \includegraphics[width=\linewidth]{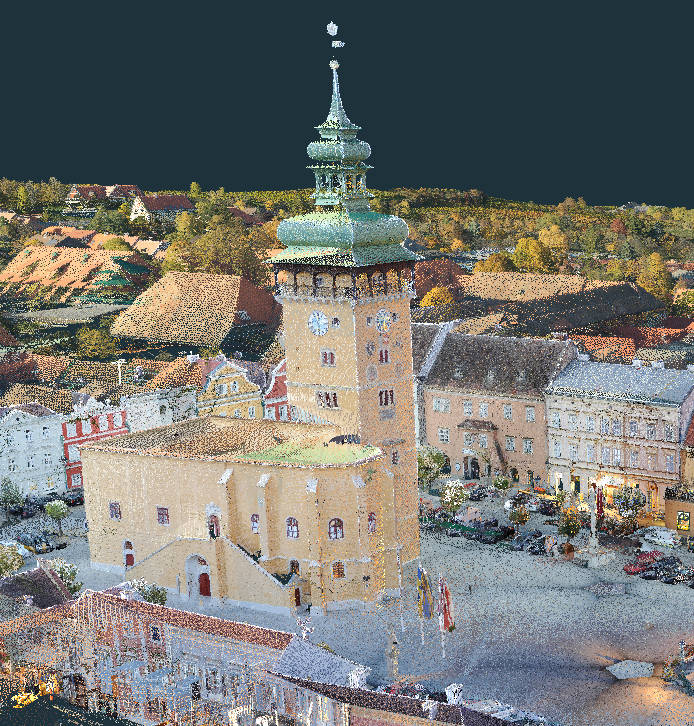}
    \caption{Retz, rendered with \enum{GL\_POINTS}}
    \end{subfigure}
    \hfill
    \begin{subfigure}{0.49\linewidth}
    \includegraphics[width=\linewidth]{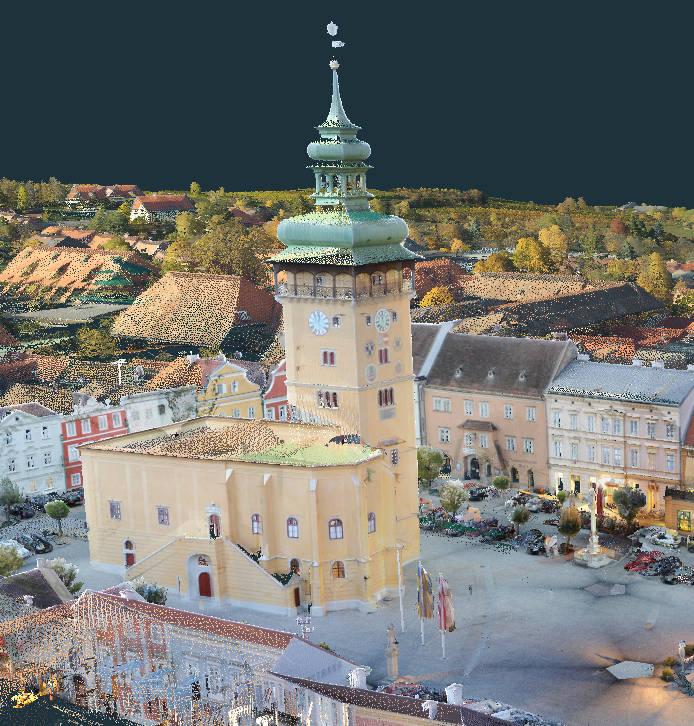}
    \caption{Retz, rendered with \textbf{HQS1R}}
    \end{subfigure}
\end{figure}

% \begin{figure}[h!]
%     \centering
%     \begin{subfigure}{0.49\linewidth}
%     \includegraphics[trim=370 180 370 60, clip,width=\linewidth]{images/im6}
%     \caption{Retz, rendered with \enum{GL\_POINTS}}
%     \end{subfigure}
%     \hfill
%     \begin{subfigure}{0.49\linewidth}
%     \includegraphics[trim=370 180 370 60, clip,width=\linewidth]{images/im5}
%     \caption{Retz, rendered with \textbf{HQS1R}}
%     \end{subfigure}
% \end{figure}

\begin{figure}[h!]
    \centering
    \begin{subfigure}{0.49\linewidth}
    \includegraphics[trim=370 180 370 60, clip,width=\linewidth]{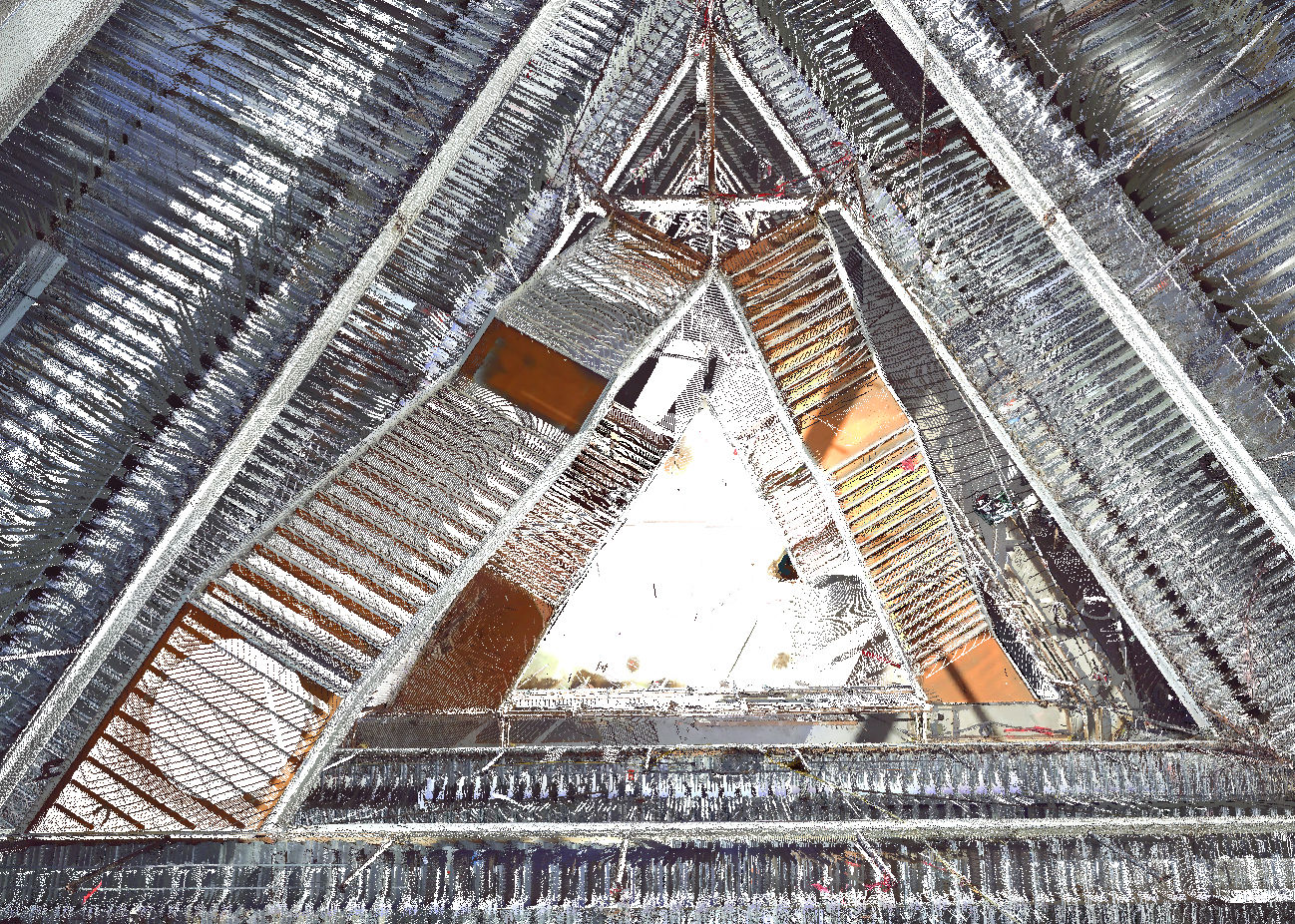}
    \caption{Endeavor staircase, rendered with \enum{GL\_POINTS}}
    \end{subfigure}
    \hfill
    \begin{subfigure}{0.49\linewidth}
    \includegraphics[trim=370 180 370 60, clip,width=\linewidth]{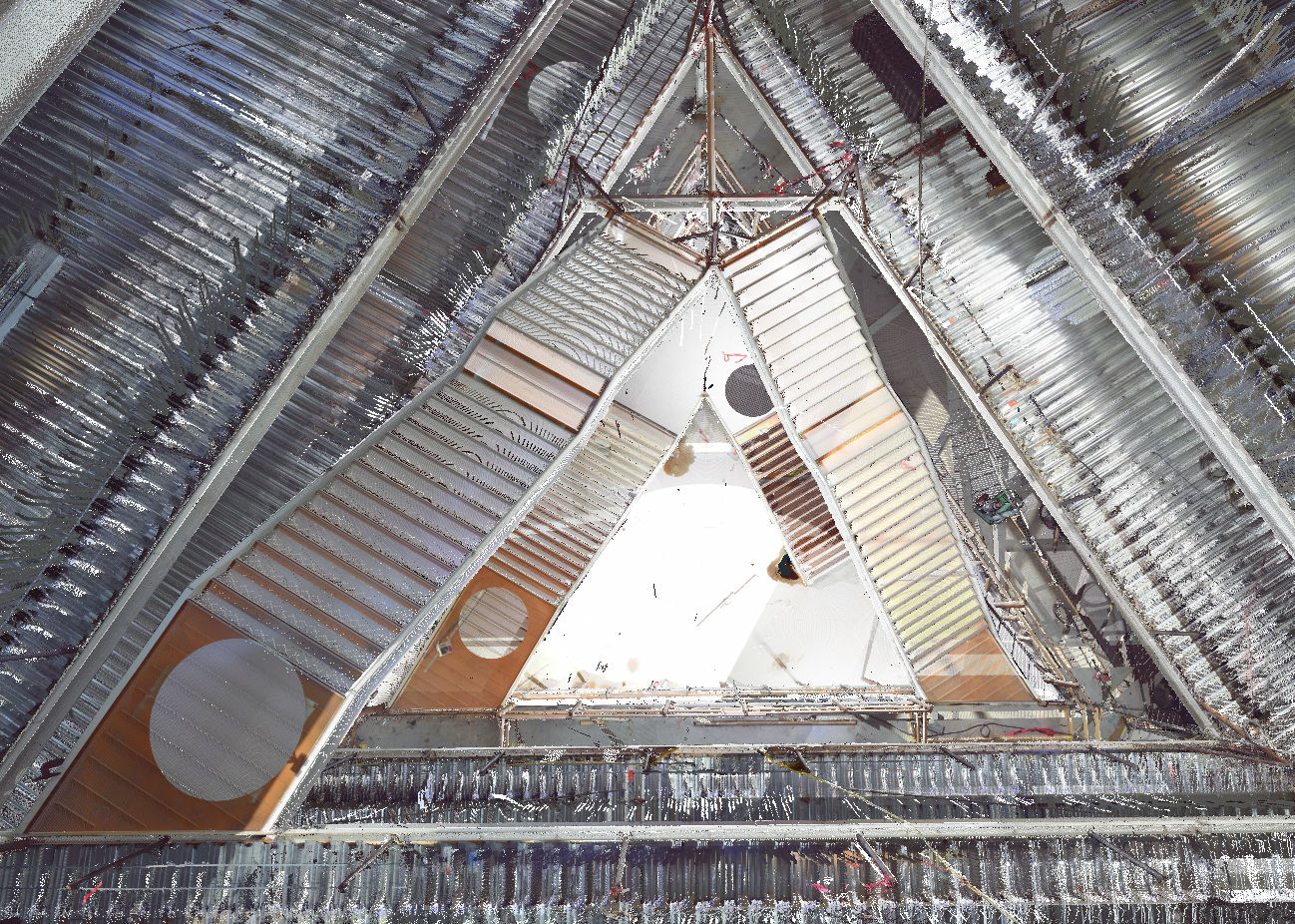}
    \caption{Endeavor staircase, rendered with \textbf{HQS1R}}
    \end{subfigure}
\end{figure}

\end{document}